\documentclass[conference]{IEEEtran}

\usepackage[fleqn]{amsmath}
\usepackage{amssymb}
\usepackage{listings} 
\usepackage{url}
\usepackage{graphicx}
\usepackage{nameref}
\usepackage{multirow}
\usepackage[inline]{enumitem}
\usepackage{array}
\usepackage{multicol}
\usepackage[nospace,noadjust]{cite}
\usepackage{diagbox}
\usepackage{color}
\usepackage{listings}
\usepackage[table,x11names,svgnames]{xcolor}
\usepackage{wrapfig}
\usepackage{lscape}
\usepackage{rotating}
\usepackage{epstopdf}
\usepackage{mathtools}
\usepackage{subcaption}
\usepackage{makecell}
\usepackage{hyperref}
\usepackage{hypcap} % fix the links
\usepackage[export]{adjustbox}
\usepackage{enumitem}

\begin{document}

\bstctlcite{IEEEexample:BSTcontrol}

\title{The Co-Evolution of Test Maintenance and Code Maintenance through the lens of Fine-Grained Semantic Changes}

\author{\IEEEauthorblockN{Stanislav Levin}
\IEEEauthorblockA{The Blavatnik School of Computer Science\\
 Tel Aviv University\\
Tel-Aviv, Israel\\
stanisl@post.tau.ac.il}
\and
\IEEEauthorblockN{Amiram Yehudai}
\IEEEauthorblockA{The Blavatnik School of Computer Science\\
 Tel Aviv University\\
Tel-Aviv, Israel\\
amiramy@tau.ac.il}
}

\maketitle

\begin{abstract}

Automatic testing is a widely adopted technique for improving software quality. Software developers add, remove and update test methods and test classes as part of the software development process as well as during the evolution phase, following the initial release.
In this work we conduct a large scale study of 61 popular open source projects and report the relationships we have established between test maintenance, production code maintenance, and semantic changes  (e.g, statement added, method removed, etc.). performed in developers' commits. 

We build predictive models, and show that the number of tests in a software project can be well predicted by employing code maintenance profiles (i.e., how many commits were performed in each of the maintenance activities: corrective, perfective, adaptive).
Our findings also reveal that more often than not, developers perform code fixes without performing complementary test maintenance in the same commit (e.g., update an existing test or add a new one). When developers do perform test maintenance, it is likely to be affected by the semantic changes they perform as part of their commit.

Our work is based on studying 61 popular open source projects, 
comprised of over 240,000 commits consisting of over 16,000,000 semantic change type instances, 
performed by over 4,000 software engineers.

\end{abstract}


\section{Introduction}

Automated testing, and automatic unit tests \cite{hamill2004unit} in particular, is a popular technique for improving software quality.
Our work \cite{levinIcsme2016,levinICSME2017_1} showed that semantic changes \footnote{A.K.A, fine-grained source code changes \cite{fluri2007change, marsavina2014studying}.}, such as method removed, field added, are statistically significant in the context of software code maintenance. Moreover, semantic changes can be used to effectively classify commits into maintenance activities (as defined by Mockus et al. \cite{mockus2000identifying}). 
These studies may indicate that semantic changes can also be useful in the context of test maintenance, and particularly, in exploring the co-evolution of test maintenance and production code maintenance, using the semantic changes as the lowest common denominator.

The field of software evolution research can be classified into two groups, the first considers the term evolution as a verb while the second as a noun \cite{lehman2000evolution}. 
\begin{description}
    \item [The verbal view] research is concerned with the question of ``how'', and focuses on means, processes, activities, languages, methods, tools required to effectively and reliably evolve a software system.
    \item [The nounal view] research is concerned with the question of ``what'' and investigates the nature of software evolution, i.e., the phenomenon, and focuses on the nature of evolution, its causes, properties, characteristics, consequences, impact, management and control \cite{lehman2000evolution,lehman2003software}.
\end{description}
Lehman et. al. \cite{lehman2000evolution,lehman2003software} suggest that both views are mutually supportive. Moreover, it is suggested that the verbal view research will benefit from progress made in studying the nounal view, and both are required if the community is to make progress in mastering software evolution.

Our intuition is that exploring the co-evolution of test maintenance and production code maintenance may benefit both the verbal and nounal views.
In terms of the verbal view, our study may help managers and practitioners reduce costs by improving the quality of their code artifacts.
For example, if certain semantic changes tend to be under-tested, it could be beneficial to design tools that can detect such changes and act upon them, whether by prompting the developers to take measures or by automatically performing scripted mitigation. 
% The question of what measures are to be taken in such case is also a non trivial one, deserving a study of its own. We can however, speculate that even a simple pop-up message saying "Your commit contains changes that are likely to involve test maintenance, have you updated the tests accordingly?" may be of value. Of course, we can imagine this becoming an automatic check that is capable of verifying whether a developer did indeed run tests before trying to commit, and only alerting in case (s)he did not. Such checks are expected to be much cheaper than performing test selection algorithms, and can serve as a first layer in an N-layer inspection mechanism.
% To reduce noise and prevent such a check from constantly alerting (in which case over time it will be simply ignored by the developers), only those changes that have the highest correlations may be considered.

In terms of the nounal view, in order to be able to design and implement such tools, we must first better understand the nature of the relationship between test evolution and production code evolution in general, and between test maintenance activities and code maintenance activities in particular. 
This work is also motivated by the rapidly growing number of production grade open source projects
hosted on a web based platform such as GitHub \cite{gitHub}, BitBucket \cite{bitbucket} and others \cite{sourceForge,codeplex,googleCode}. All make a great source of publicly available, free, and high quality source code corpora which was not available in the earlier stages of studying software and test evolution. The absence of such corpora, often resulted in the models for growth dynamics being relatively simplistic \cite{lehman2003software}. Moreover, along with the large source code corpora now available, progress has been made in the Big Data ecosystem \cite{diebold2012origin}, bringing software tools capable of processing extremely large data volumes to the masses.
The combination of the two has created unprecedented opportunities to collect and process an enormous volume of source code \cite{levinIcsme2016}, and provide insights that were previously exponentially harder, or even impossible to obtain \cite{raychev2015predicting}.

Our study concentrates on the following research questions:
\begin{enumerate}[leftmargin=*,labelindent=16pt, label={\textbf{RQ. \arabic*.}}]
    \item How does (production) code maintenance relate to projects' tests count?
    \item How often is test maintenance performed as part of (production) code maintenance?
    \item How do semantic changes performed in (production) code maintenance relate to test maintenance activities?    
\end{enumerate}

\section{Data Collection}

We harvest software code repositories from GitHub (\cite{gitHub}), a popular repository hosting platform with rich query options. Candidate repositories were selected according to the following criteria, which we designed to target data-rich repositories:
\begin{itemize}
    \item Had over 900 Java commits (i.e., commits where Java files were changed)
    \item Were created before 2015-01-01  (i.e., these repositories had been around for a while)
    \item Had size over 2MB (i.e. these repositories are of considerable size)
    \item Had more than 100 stars (i.e. more than 100 users had ``liked'' these repositories)
    \item Had more than 60 forks (i.e., more than 60 users had ``copied'' these repositories for their own use)
    \item Had their code updated since 2016-01-01 (i.e., these repositories were active)    
\end{itemize}

To perform the data collection and processing tasks we use a designated VCS mining platform we have built on top of Spark \cite{zaharia2010spark, sparkSite}, a state of the art framework for large data processing.
Our final dataset consisted of 61 projects {\renewcommand{\footnotesize}{\scriptsize}\footnote{https://github.com/staslev/paper-resources/blob/icsme-2017/The-Co-Evolution-of-Test-Maintenance-and-Code-Maintenance-through-the-lens-of-Fine-Grained-Semantic-Changes/studied-repos.md}} from various domains, such as IDEs, programming languages (that were implemented in Java), distributed databases and storage platforms, integration frameworks and more (see summary statistics in table \ref{tab:datasetStats}). This dataset included a total of 242,567 commits, 4,259 developers and 16,161,680 semantic changes.

\begin{table}[ht]
  \centering 
  \caption{Project metrics statistics} 
  \label{tab:datasetStats}   
  \normalsize
  \renewcommand{\arraystretch}{1.5}  
  \resizebox{\linewidth}{!}{
  \begin{tabular}{|p{0.19\linewidth}|c|c|c|c|c|c|} 
    \cline{2-6}
    \multicolumn{1}{c|}{} &\cellcolor{lightgray} Min & \cellcolor{lightgray} 1-Q. & \cellcolor{lightgray} Median & \cellcolor{lightgray} Mean & \cellcolor{lightgray} 3-Q. &  \cellcolor{lightgray}Max \\
    \hline
    \cellcolor{lightgray} LOC (Lines Of Code) & 13,710 &   58,010 &  123,500 &  213,100 &  334,000 &  993,300 \\
    \hline
    \cellcolor{lightgray} Developers & 10 &  27 &  52 &  70 &  95 & 295 \\
    \hline
    \cellcolor{lightgray} Age (days) & 412 &  1,600 &   2,527  &  2,562 &   3,521 &   6,792 \\    
    \hline
    \cellcolor{lightgray} Commits Analysed & 1,007 &   1,493 &  2,539 & 3,899 & 5,844 & 20,180 \\
    \hline
    \cellcolor{lightgray} Semantic Changes Extracted & 26,360 &  78,530 & 173,300 & 264,900 & 325,500 & 1,166,000 \\
    \hline
\end{tabular}}
\end{table}

\subsection{Distilling Semantic Changes}

After downloading (cloning) the repositories from GitHub, for each repository $r$ where $1\leq r \leq 61$ we created a series of patch files $\{p_{i}^{r}\}_{i=1}^{N_r}$, where $N_r$ is the latest revision number for repository $r$. Each patch file $p_{i}^{r}$ was responsible for transforming repository $r$ from revision $r_{i-1}$ to revision $r_{i}$, where $r_0$ is the empty repository. By initially setting repository $r$ to revision $1$ (i.e. the initial revision) and then applying all patches ${\{p_{i}^{r}\}}_{i=2}^{N_r}$ in a sequential manner, the revision history for that repository was essentially replayed. Conceptually, this was equivalent to the case of all developers performing their commits sequentially one by one according to their chronological order.

To distill semantic change types as per the taxonomy defined by Fluri et al., we repeatedly applied a customized version of the ChangeDistiller tool (\cite{gall2009change, fluri2008discovering, martinez2013automatically,fluri2007change}) on every two consecutive revisions of every Java file in every repository we had selected to be part of the dataset. We had to perform some modifications to the original ChangeDistiller tool since we encountered use cases where the distilled change list was incomplete. In particular, the addition and removal of classes did not produce changes corresponding to all the internal methods being added or removed as well. Since this scenario is crucial for the extraction of test maintenance activities (including test method addition and removal as a result of a test class addition or removal), we enhanced the original ChangeDistiller with this feature.

\subsection{Classifying Commits Into Maintenance Activities}\label{sec:maintActivities}

Three main classification categories for maintenance activities in software projects were identified by Mockus et al. \cite{mockus2000identifying}:
\begin{itemize}
    \item Corrective: fixing faults, functional and non-functional.
    \item Perfective: improving the system and its design.
    \item Adaptive: introducing new features into the system.
\end{itemize}
In our work on commit classification into maintenance activities \cite{levinICSME2017_1} we suggested a model for cross-project commit classification that was able to achieve an accuracy of 76\%.
The suggested technique consists of the following steps:
\begin{enumerate}
    \item Manually classify the maintenance activities for a ground truth set
    \item For each maintenance activity (``Corrective'', ``Perfective'' or ``Adaptive''), perform a word frequency analysis using the ground truth set commits' comments in order to obtain the 10 most frequent words for each maintenance activity
    \item Distill the semantic changes from the commits in the ground truth set,
    and perform a frequency analysis
    \item Use the ground truth set to train a RandomForest \cite{ho1998random, breiman2001random} based classification model, where features are the combination of commits' word frequencies and semantic change type frequencies obtained from in previous steps
\end{enumerate}
For the purpose of this work, we use the classification model we have devised in \cite{levinICSME2017_1}.

\subsection{Detecting Test Maintenance}\label{sec:testActivities}
\label{sec:testRelatedOperations}

In the scope of this work, we consider a class to be a \textit{test class} (a.k.a test suite) either if its name starts with the word ``Test'', or, if it ends with the words ``Test'', or ``Tests'', or ``TestCase''.
We consider a method to be a \textit{test method} (a.k.a test case) if it has a ``Test'' Java annotation, OR starts with the word ``test'' and resides inside a test class. These heuristics are popular both in the software industry \cite{mavenTests} and academia \cite{zaidman2011studying}.

We use semantic changes to detect test maintenance by inspecting the following semantic change types:
\begin{itemize}[topsep=0pt]
    \item {\small ``ADDITIONAL\_FUNCTIONALITY''} - the addition of a new method
    \item {\small ``REMOVED\_FUNCTIONALITY''} - the removal of an existing method
    \item {\small ``ADDITIONAL\_CLASS''} - the addition of a new class
    \item {\small ``REMOVED\_CLASS''} - the removal of an existing class
    \item Other change type (see \cite{fluri2006classifying} for the full list) - if the parent entity qualifies as a test method or a test class, the original entity is classified as a test method update or a test class update, respectively.
\end{itemize}
We detect test method/class activity as part of a commit by inspecting its semantic change types and the corresponding element names.

To capture the test maintenance activities, given a commit $c$ we define the following metrics:
\begin{itemize}
\setlength{\itemsep}{5pt}    
    \item $\operatorname{Test^{Method}_{A}}(c)$, number of test methods added in $c$.
    \item $\operatorname{Test^{Method}_{R}}(c)$, number of test methods removed in $c$.
    \item $\operatorname{Test^{Method}_{U}}(c)$, number of test methods updated in $c$.
    \item $\operatorname{Test^{Class}_{A}}(c)$, number of test classes added in $c$.
    \item $\operatorname{Test^{Class}_{R}}(c)$, number of test classes removed in $c$.
    \item $\operatorname{Test^{Class}_{U}}(c)$, number of changes inside a test class, but outside any of the test methods in that class, as part of $c$, e.g., setUp, tearDown \cite{junitTestFixture}, and other helper methods.    
    \item $\operatorname{TestMaintenance}$, the total number of test activities performed as part of a given commit.
        \begin{multline*}
        \operatorname{TestMaintenance}(commit) := \\ \sum\limits_{\substack{\operatorname{scope}\in\{\operatorname{Method,Class}\}\\
        \operatorname{activity}\in\{\operatorname{A,R,U}\}}}\operatorname{Test_{activity}^{scope}}
        \end{multline*}
\end{itemize}

\section{Statistical Methods}

We use regression models to study the relationships between a set of predictors and an outcome variable. In particular, we use generalized linear modeling  (GLM) \cite{mccullagh1989generalized} for count and for logistic regressions to explore the effects of maintenance activities and semantic change types on test method and class counts, and test maintenance activities.

For count regression models, we report the statistically significant predictors, and use their coefficients to analyse their effect on the outcome variable. We then build predictive models by keeping only the most significant predictors. Finally, we use analysis of variance (ANOVA \cite{chambers1991statistical}) to establish the magnitude of predictors' effects by observing the reduction in the residual deviance associated with the variable’s effect in the model. We evaluate the predictive models by splitting our dataset into a training and a validation datasets, the former is used for training the models, and the latter for evaluating their predictive powers. We report the $\operatorname{p-value}$ as a measure of goodness of fit. A high $\operatorname{p-value}$ indicates a lack of evidence to support the hypothesis that the observed counts do not match the expected counts, implying a good fit. 

In case of logistic regression models, we report both statistically significant predictors, and the odds ratio along with their 95\% confidence level intervals. The odds in favor of an outcome $A$ is the ratio of the probability of an outcome A to occur and the probability of the complement of A (i.e., that A will not occur) and is
defined as $\frac{P(A)}{P(A^c)}$ \cite{fulton2012confusion}.
The odds ratio represents the odds that an outcome will occur given an exposure to a particular effect, compared to the odds of the outcome occurring in the absence of that exposure.
Odds ratios are used to compare the relative odds of the occurrence of the outcome of interest, given exposure to the variable(s) of interest. 
Odds ratio greater than 1 indicates an increase in the odds in favor of the outcome, while odds ratio less than 1 indicates a decrease in the odds in favor of the outcome. Odds ratio equals 1 indicates no effect on the odds of the outcome given a particular exposure.

We use the R statistical environment \cite{R} for statistical computations, where we extensively use the R caret package \cite{rCaret} for the purpose of model training and evaluation.

\section{Production (code) Maintenance and Test Maintenance}

\subsection{\textbf{RQ. 1:} How does (production) code maintenance relate to projects' tests count?}
\label{sec:testCountRelate}

The statistics for test method and test class counts in the projects we studied can be found in table \ref{tab:projectTestMethodsClasses}. Since these counts vary greatly in absolute numbers and are highly dependent on the size of the project, we report them  per 1000 LOC to give a standardized perspective. 
We also compute the average number of test methods in a test class, which stands at 4.636.

\begin{table}[ht]
  \centering 
  \caption{Test method and class metrics statistics} 
  \label{tab:projectTestMethodsClasses} 
  \renewcommand{\arraystretch}{2}
  \begin{tabular}{|l|p{0.04\linewidth}|p{0.07\linewidth}|p{0.09\linewidth}|p{0.07\linewidth}|p{0.07\linewidth}|p{0.07\linewidth}|} 
    \cline{2-6}
     \multicolumn{1}{c|}{} & \cellcolor{lightgray} Min & \cellcolor{lightgray} 1-Q. & \cellcolor{lightgray} Median & \cellcolor{lightgray} Mean & \cellcolor{lightgray} 3-Q. &  \cellcolor{lightgray} Max \\
    \hline
    \cellcolor{lightgray} $\frac{\operatorname{Test}^{\operatorname{Methods}}}{1000_{ LOC}}$ & 0 & 4.68  & 9.38 & 11.21 & 14.73 & 52.62 \\
    \hline
    \cellcolor{lightgray} $\frac{\operatorname{Test}^{\operatorname{Classes}}}{1000_{ LOC}}$ & 0 & 0.898 & 2.126 & 2.503 & 3.677 & 10.51  \\
    \hline
\end{tabular}
\end{table}

% \begin{figure}[ht]    
%     %\captionsetup{belowskip=12pt}
%     \centering
%     \caption{Number of test classes and methods per 1000 LOC}
%     \label{fig:testsPerLoc}    
%     \includegraphics[width=0.8\linewidth,height=0.8\linewidth]{figures/tests-per-loc}        
%     \vspace{-2em}
% \end{figure}

To better understand the relations between the number of tests (methods and classes), and maintenance activities we devise GLMs of the form: $${\operatorname{Test}^M(prj) = \operatorname{C^M} + \sum\limits_{i=1}^{|\operatorname{Predictors}|}(\operatorname{coeff}^M_i * \operatorname{predictor}^M_i(prj))}$$ where $M$ is the test metric we model, i.e., ${M \in \{\operatorname{Methods}, \operatorname{Classes}\}}$, $\operatorname{Predictors}$ is the predictors set, $\operatorname{coeff}^M_i$ are the predictor coefficients, $\operatorname{predictor}^M_i(prj)$ are predictor values, and $\operatorname{C^M}$ is the model constant.
The corresponding models for $\operatorname{Test}^{\operatorname{Methods}}$ and $\operatorname{Test}^{\operatorname{Classes}}$ can be found in table \ref{mod:prjTests}.
All predictors were log transformed to alleviate skewed data, a common practise when dealing with software metrics \cite{shihab2012exploration, camargo2009towards}.
Statistically significant predictors of interest are highlighted in lime-green, and the standard error is reported in parenthesis below the estimated coefficients. 
In addition to the variables we are directly interested in, such as the $\operatorname{log(corrective)}$, $\operatorname{log(perfective)}$ and $\operatorname{log(adaptive)}$ we also use $\operatorname{log(LOC)}$, $\operatorname{log(age)}$ and $\operatorname{log(developers)}$ as control variables, in order to reduce the effect of lurking variables which correlate both with the predictors and the predicted (outcome) variable. Control variables are highlighted in light-bisque. The models in table \ref{mod:prjTests} were devised using all 61 projects in our dataset.

% Table created by stargazer v.5.2 by Marek Hlavac, Harvard University. E-mail: hlavac at fas.harvard.edu
% Date and time: Mon, Jan 30, 2017 - 19:53:17
\begin{table}[htbp] \centering 
  \caption{Negative Binomial GLM (\cite{venables2013modern}) for test method and test class counts} 
  \label{mod:prjTests} 
\begin{tabular}{@{\extracolsep{5pt}}lcc} 
\\[-1.8ex]\hline 
\hline \\[-1.8ex] 
 & \multicolumn{2}{c}{\textit{Dependent variable:}} \\ 
\cline{2-3} 
Predictor & $\operatorname{Test}^{\operatorname{Methods}}$ & $\operatorname{Test}^{\operatorname{Classes}}$  \\ 
\hline \\[-1.8ex] 
 \rowcolor{lime} log(corrective) & $-$1.696$^{***}$ & $-$1.351$^{***}$ \\ 
  & (0.314) & (0.285) \\ 
  & & \\ 
 \rowcolor{lime} log(perfective) & 1.621$^{***}$ & 1.583$^{***}$ \\ 
  & (0.397) & (0.358) \\ 
  & & \\ 
 log(adaptive) & $-$0.247 & $-$0.173 \\ 
  & (0.366) & (0.329) \\ 
  & & \\ 
 \rowcolor{Bisque1} log(developers) & 0.318$^{*}$ & 0.105 \\ 
  & (0.182) & (0.163) \\ 
  & & \\ 
 \rowcolor{Bisque1} log(LOC) & 1.189$^{***}$ & 1.053$^{***}$ \\ 
  & (0.171) & (0.154) \\ 
  & & \\ 
 \rowcolor{Bisque1} log(age) & 0.770$^{***}$ & 0.686$^{***}$ \\ 
  & (0.205) & (0.185) \\ 
  & & \\ 
 Constant & $-$12.326$^{***}$ & $-$13.289$^{***}$ \\ 
  & (1.873) & (1.702) \\ 
  & & \\ 
\hline \\[-1.8ex] 
Observations & 61 & 61 \\ 
\hline 
\hline \\[-1.8ex] 
\textit{Note:}  & \multicolumn{2}{r}{$^{*}$p$<$0.1; $^{**}$p$<$0.05; $^{***}$p$<$0.01} \\ 
\end{tabular} 
\end{table} 

\subsection{Analysing the count regression model}
\label{sec:rq1Discussion}

Greater project size (in LOC) and/or age (in days) are likely to boost the the project's expected number of tests (method and class), as indicated by the positive coefficients of the $\operatorname{log(LOC)}$, $\operatorname{log(age)}$ predictors.
These results align well with our intuition according to which mature projects, that are larger in size (LOC wise) and have been developed over long periods of time, are likely to contain more tests.

Higher number of perfective commits, i.e., commits that aim to improve system design and code structure, is also likely to boost the expected number of tests (methods and classes) in a project.
Higher numbers of corrective commits on the other hand, i.e., commits that aim to fix faults, is likely to associate with projects with less tests (methods and classes), as
indicated by the negative coefficient of the $\operatorname{log(corrective)}$ predictor for both $\operatorname{Test^{Methods}}$ and $\operatorname{Test^{Classes}}$.
In spite of the fact that regression models do not provide means to ascertain causality, the negative coefficient of corrective commits on tests (i.e., both methods and classes) is worth considering. Potentially, one could argue that projects with tests may only need little corrective activity due to the high quality of the codebase. The opposite direction, may imply that corrective activity may be required when the test count of a project is low, and the codebase's quality is poor. It is also possible, that test counts and corrective commits do not have a cause and effect relationship at all, in which case they just tend to happen together and are connected via a lurking variable.
The effect of adaptive commits on test counts is inconclusive as their coefficients did not demonstrate a statistical significance.

\begin{figure}[ht]    
    %\captionsetup{belowskip=12pt}
    \caption{$\operatorname{Test}^{\operatorname{Methods}}$, predicted vs. actual}
    \label{fig:modTestMethodValidation}    
    \includegraphics[width=\linewidth]{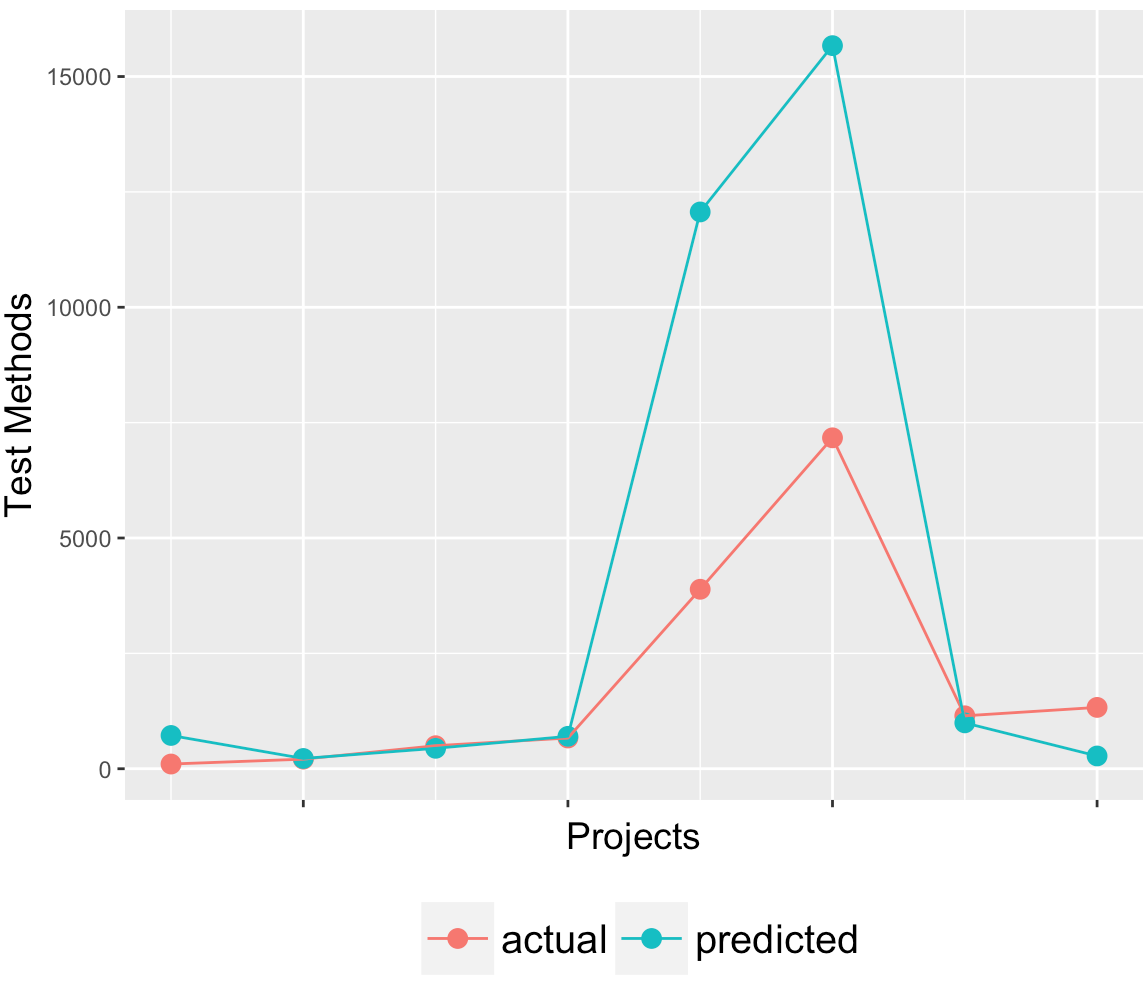}        
    %\vspace{1em}%
\end{figure}

\begin{figure}[ht]    
    %\captionsetup{belowskip=12pt}
    \caption{$\operatorname{Test}^{\operatorname{Classes}}$, predicted vs. actual}
    \label{fig:modTestClassValidation}    
    \includegraphics[width=\linewidth]{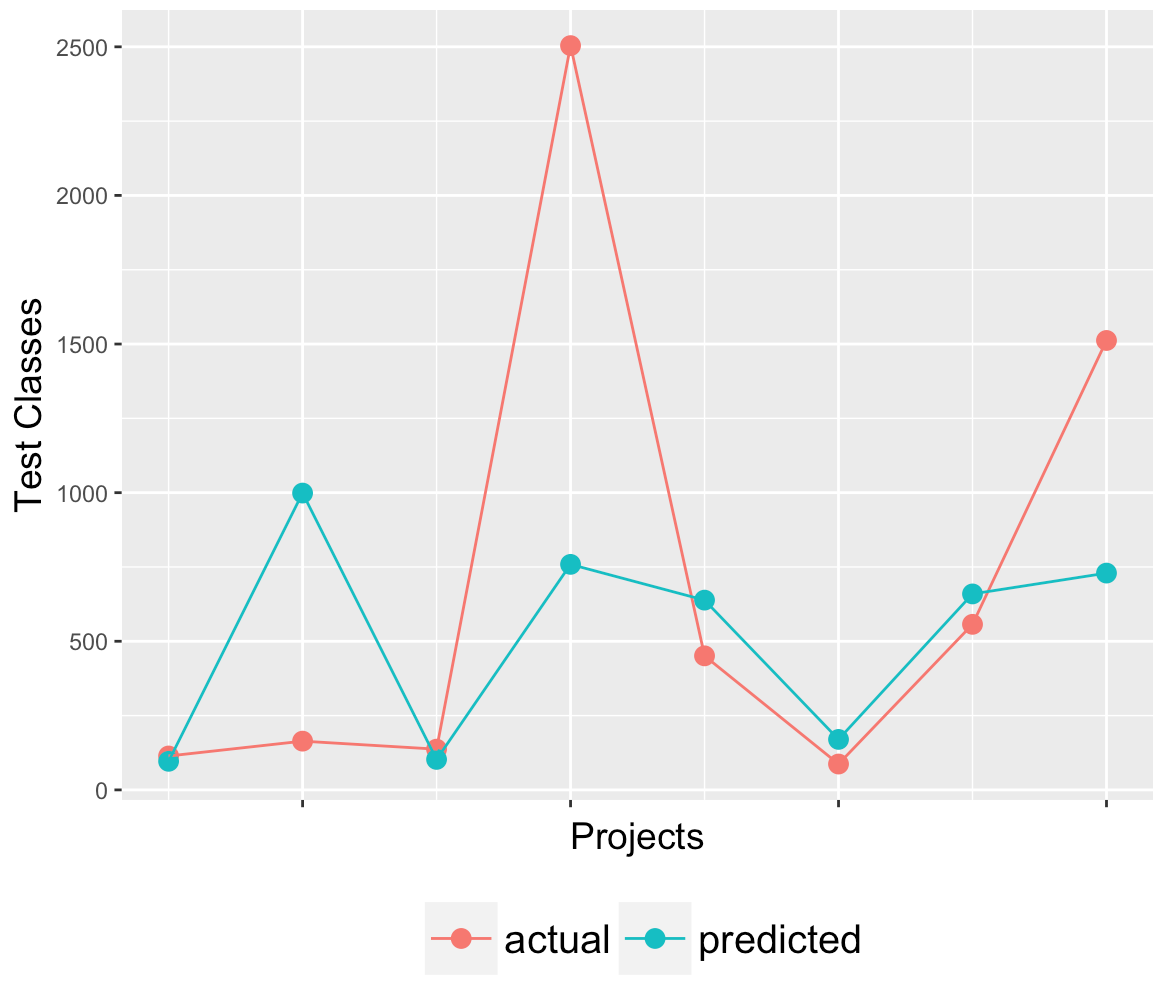}        
    %\vspace{1em}%
\end{figure}

In addition to establishing the correlations between maintenance activities and test counts, our models can also be used to predict the latter using the former. In order to improve prediction results we optimise the models by keeping only the most statistically significant predictors (i.e. with $\operatorname{p-value} < 0.01$): $\operatorname{log(corrective)}$, $\operatorname{log(perfective)}$, $\operatorname{log(LOC)}$ and $\operatorname{log(age)}$. 
The resulting models were then trained using 53 randomly selected projects (our of 61) from our dataset, and tested using the remaining 8 ($\sim$13\% of the entire dataset).
Figure \ref{fig:modTestMethodValidation} and figure \ref{fig:modTestClassValidation} chart the predicted vs. the actual values of $\operatorname{Test}^{\operatorname{Methods}}$ and $\operatorname{Test}^{\operatorname{Classes}}$, respectively. The x-axis is a running index ($1\dots8$, indicating the index of the project in our validation data set), and the y-axis is the number of tests. The red line depicts the actual values, and the turquoise line depicts the predicted ones.
Our predictions for projects' test methods and test classes show a substantial accuracy, as indicated by the $\operatorname{p-value}$, which was 0.067 and 0.071, respectively (values greater than 0.05 indicate high goodness of fit). 
The ANOVA for the predictive models ($\operatorname{Test^{Methods}}$ and $\operatorname{Test^{Classes}}$), which we use to establish the magnitude of predictors' effects, can be found in table \ref{tab:anovaTestMethods} and \ref{tab:anovaTestClasses}, respectively.
Each row indicates the reduction in the residual deviance and degrees of freedom induced by adding a given predictor to the model.
By observing the ``Deviance'' column of tables \ref{tab:anovaTestMethods} and \ref{tab:anovaTestClasses} we learn that both perfective and corrective have a substantial contribution to ``explaining'' the test counts (tests and methods). 
Perfective commits have the greatest effect in both the test method and test class models, while the corrective commits's effect is 52\% less in the $\operatorname{Test}^{\operatorname{Methods}}$ model and only 1\% less in the $\operatorname{Test}^{\operatorname{Classes}}$ model (see cells highlighted in yellow in table \ref{tab:anovaTestMethods} and table \ref{tab:anovaTestClasses}).
This may imply that corrective commits play a greater role in predicting test classes than test methods. Also worth noting is the LOC predictor, which demonstrated high explanatory power in both test method and test class predictive models, indicating that the size of the project does indeed have a considerable effect on the number of test methods and test classes it contains.

% latex table generated in R 3.3.1 by xtable 1.8-2 package
% Mon Jan 30 20:37:41 2017
\begin{table}[h!]
\centering
\centering
  \caption{ANOVA for $\operatorname{Test^{Methods}}$} 
  \label{tab:anovaTestMethods} 
\begin{tabular}{|p{0.17\linewidth}|p{0.03\linewidth}|p{0.1\linewidth}|p{0.1\linewidth}|p{0.14\linewidth}|c|}
  \cline{2-5}
  \multicolumn{1}{c|}{} & \cellcolor{lightgray} Df. & \cellcolor{lightgray} Deviance & \cellcolor{lightgray} Residual Df. & \cellcolor{lightgray} Residual Deviance & \cellcolor{lightgray} Pr($>$Chi) \\ 
  \hline
 \cellcolor{lightgray} NULL &  &  & 52 & 155.57 &  \\ 
\hline
\cellcolor{lightgray} log(corrective) & 1 & \cellcolor{yellow} 18.78 & 51 & 136.79 & 0.0000 \\ 
  \hline
  \cellcolor{lightgray} log(perfective) & 1 & \cellcolor{yellow} 38.76 & 50 & 98.03 & 0.0000 \\ 
  \hline
  \cellcolor{lightgray} log(LOC) & 1 &  \cellcolor{Bisque1} 30.74 & 49 & 67.29 & 0.0000 \\ 
  \hline
  \cellcolor{lightgray} log(age) & 1 & 3.93 & 48 & 63.36 & 0.0473 \\ 
   \hline
\end{tabular}
\end{table}

% latex table generated in R 3.3.1 by xtable 1.8-2 package
% Mon Jan 30 20:17:27 2017
\begin{table}[h!]
\centering
  \caption{ANOVA for $\operatorname{Test^{Classes}}$} 
  \label{tab:anovaTestClasses} 
\begin{tabular}{|p{0.17\linewidth}|p{0.03\linewidth}|p{0.1\linewidth}|p{0.1\linewidth}|p{0.14\linewidth}|c|}
  \cline{2-5}
  \multicolumn{1}{c|}{} & \cellcolor{lightgray} Df. & \cellcolor{lightgray} Deviance & \cellcolor{lightgray} Residual Df. & \cellcolor{lightgray} Residual Deviance & \cellcolor{lightgray} Pr($>$Chi) \\ 
  \hline
\cellcolor{lightgray} NULL &  &  & 52 & 194.16 &  \\ 
\hline
  \cellcolor{lightgray} log(corrective) & 1 & \cellcolor{yellow}  44.70 & 51 & 149.45 & 0.0000 \\ 
  \hline
  \cellcolor{lightgray} log(perfective) & 1 & \cellcolor{yellow}  45.05 & 50 & 104.40 & 0.0000 \\ 
  \hline
  \cellcolor{lightgray} log(LOC) & 1 & \cellcolor{Bisque1} 30.15 & 49 & 74.25 & 0.0000 \\ 
  \hline
  \cellcolor{lightgray} log(age) & 1 & 11.20 & 48 & 63.05 & 0.0008 \\ 
   \hline
\end{tabular}
\end{table}

The predictive models we devise (see table \ref{mod:prjTests}) show that different maintenance activities have different affect on test counts (method and class). Consequently, it implies that having prior knowledge of a commits class (corrective, perfective or adaptive) may contribute to predicting its affect on the overall test count (method and class) of a project. Moreover, these models demonstrate good prediction powers (see figure \ref{fig:modTestMethodValidation} and figure \ref{fig:modTestClassValidation}) which can benefit future applications such as resource allocation and project planning (see section \ref{sec:conclusions})

\subsection{\textbf{RQ. 2:} How often is test maintenance performed as part of (production) code maintenance?}
\label{sec:howOften}

We are interested in exploring whether test maintenance (addition, deletion or update of test methods or test classes) are common in the context of software projects. In particular, we wish to investigate whether test maintenance is typically performed as part of production code maintenance. In case the two tend to occur together, can test maintenance activities be attributed to any particular code maintenance activity type (corrective, perfective, adaptive)?

\begin{figure}[h!]    
    %\captionsetup{belowskip=12pt}
    \centering
    \caption{Percentage of commits that involve test maintenance, per project}
    \label{fig:commits-with-test-activities}    
    \includegraphics[width=0.5\linewidth,height=0.6\linewidth]{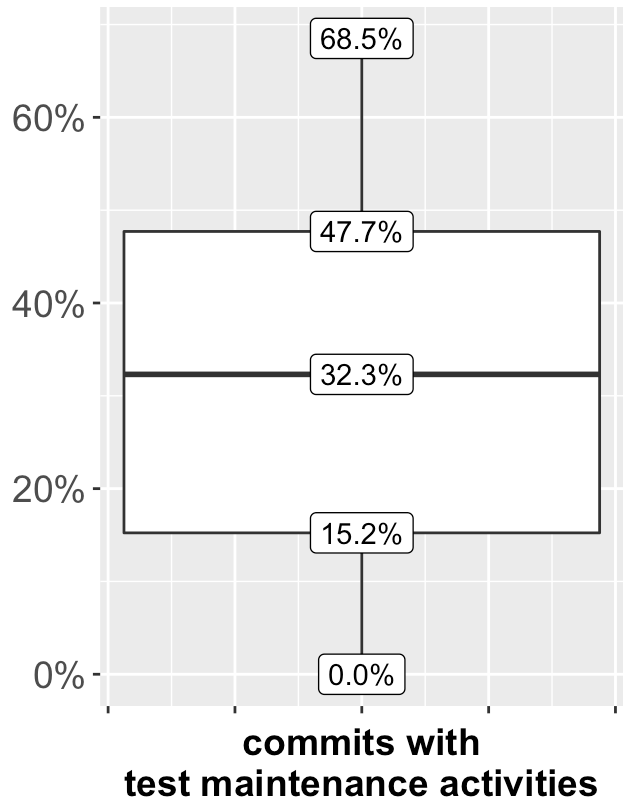}   
    %\vspace{1em}%
\end{figure}

First, we inspect how common test maintenance is within a project, without accounting for the various code maintenance activities. That is, we inspect the percentage of the commits that involve test maintenance in the scope of a project (see figure \ref{fig:commits-with-test-activities}).

The box-plot in figure \ref{fig:commits-with-test-activities} shows that proportions of test maintenance is quite different across the projects in our study, while in some projects more than half of the commits involved test maintenance, in others less than 15\%. In none of the projects, did the test maintenance occur in more than 68.5\% of the commits.
Moreover, in a few projects there were no tests at all and therefore no test maintenance took place.
This may imply testing practices remain to be determined individually in each project, rather than standardised.

\begin{figure}[h!]    
    %\captionsetup{belowskip=12pt}
    \caption{Percentage of commits that involve test maintenance within each maintenance activity type, per project}
    \label{fig:perProjectTestsByMaint}    
    \includegraphics[width=\linewidth]{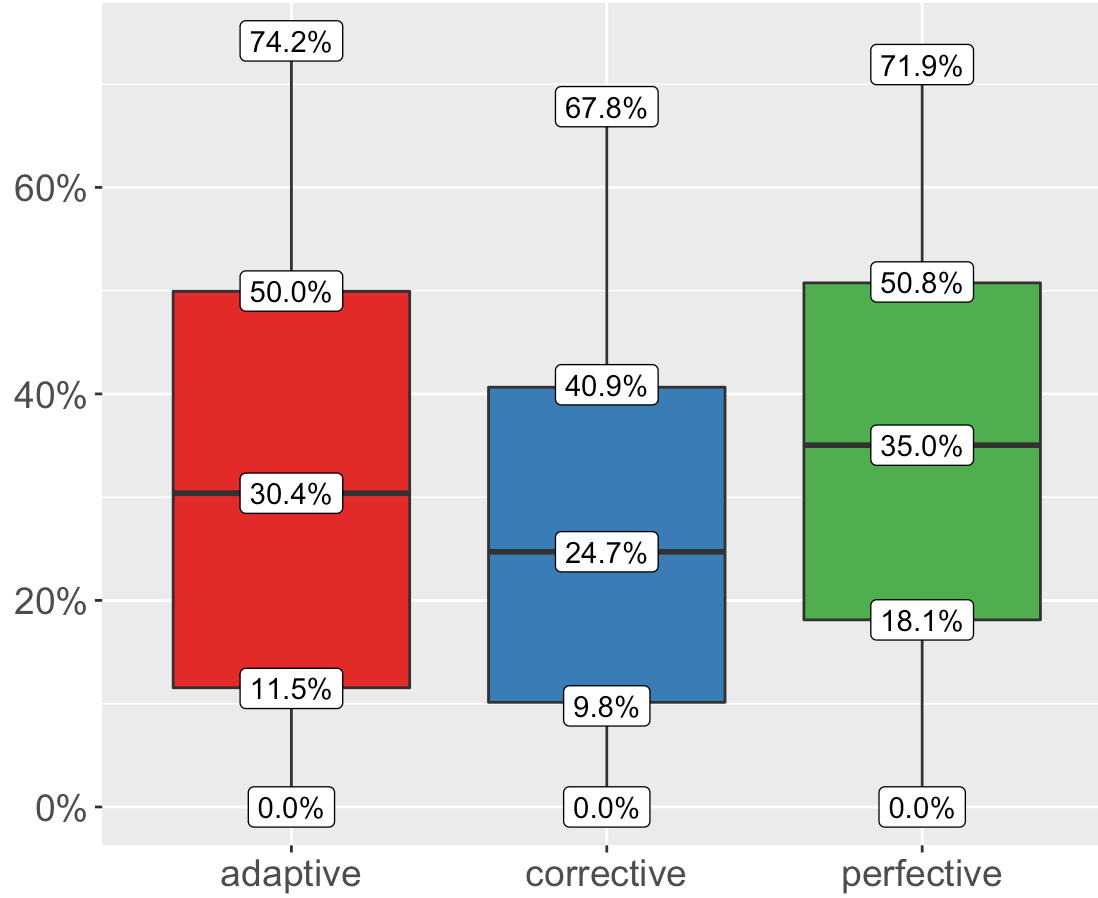}   
    %\vspace{1em}%
\end{figure}

We then analyse how common test maintenance is, within each of the code maintenance activities (corrective, perfective, adaptive), see figure \ref{fig:perProjectTestsByMaint}.

\subsection{Analysing test maintenance portions in commits}
\label{sec:rq2Discuss}

\begin{figure}[h!]    
    %\captionsetup{belowskip=12pt}
    \caption{Percentage of commits that involve test maintenance within each maintenance activity type, per developer}
    \label{fig:perProjectTestsByMaintDeveloper}    
    \includegraphics[width=\linewidth]{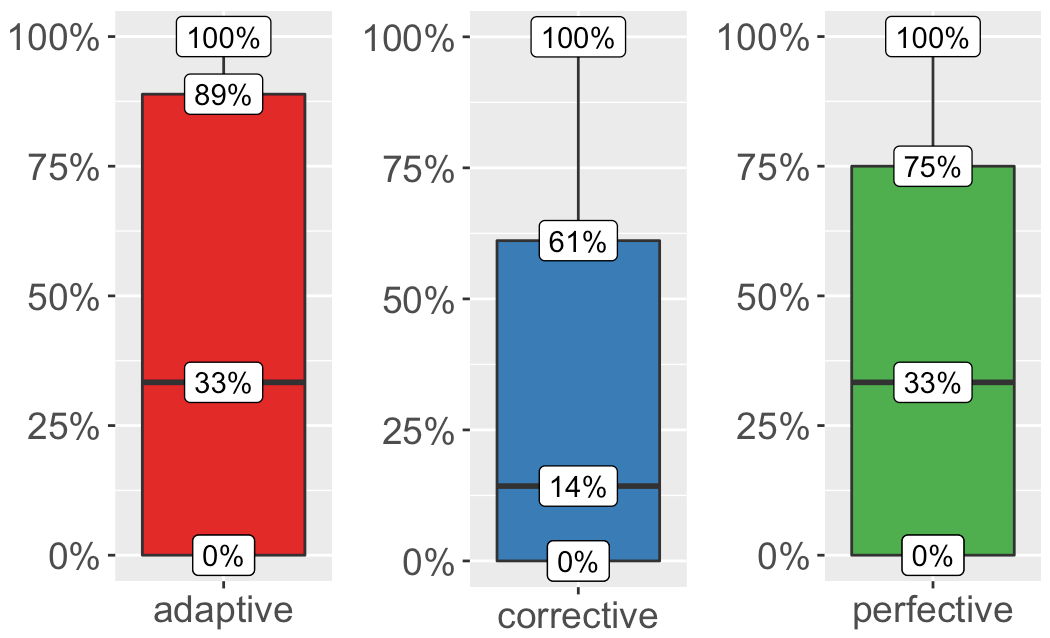}   
    %\vspace{1em}%
\end{figure}

The box-plots in figure \ref{fig:perProjectTestsByMaint} show that for half the projects (i.e., the median) in our dataset, test maintenance was present in less than 24.7\% of the corrective commits, less than 30.4\% in adaptive commits, and less then 35\% in perfective commits.
It is visually apparent that the test maintenance tends to be the least common within the corrective commits. However, to gain statistical confidence we perform similar analysis on a developer level, where more data points are available (per-developer data points vs. per-project data points).
Figure \ref{fig:perProjectTestsByMaintDeveloper} complements the per-project perspective with a per-developer-per-project one, where for each developer and a maintenance activity type ${\operatorname{MA} \in \{\operatorname{Corrective}, \operatorname{Perfective}, \operatorname{Adaptive}\}}$, we calculate the percentage of commits of type $\operatorname{MA}$ that involved test maintenance.
The Wilcoxon-Mann-Whitney (\cite{bauer1972constructing}) test confirms ($\operatorname{p-value} < 0.01$) that the percentage of test maintenance within the corrective commits is the lowest.
In addition, we can see that some developers perform test maintenance in all of their corrective commits, some perform test maintenance in all of their adaptive commits, and some perform test maintenance in all of their perfective commits. Perhaps, some perform test maintenance in all of their commits. 
The phenomena of developers performing test maintenance in 100\% of their commits in either of the maintenance categories may indeed indicate such developers are true test maintenance fans, alternatively, such cases may indicate that these particular developers perform only few commits, e.g. 1, which translates into a high relative portion (e.g. 100\%) of their commits involving test maintenance.

The lack of test maintenance in corrective commits is dominant in both project and developer oriented views. We suggest two hypotheses that may account for this phenomena:
\begin{enumerate}
    \item a developer commits a change that breaks one or more tests and thus also breaks the build performed by the Continuous Integration (CI) system \cite{fowler2006continuous, jenkins, teamCity}. Now, she must fix the breakage by performing further changes. In this case, the subsequent commit, which would be considered a corrective one, will not involve a test maintenance activity, since the developer is fixing a bug caught by a test.
    \item a developer performs a fix for a bug she has been made aware of, which is not covered by existing tests. The developer completes the fix, but does not provide an automatically runnable proof-of-fix, i.e., she does not add a new test neither does she update an exiting one, so that it will fail in case of regression.
\end{enumerate}
Neither of these scenarios is one that adheres to what may be considered best practices.
In case of a broken build, we would have expected developers to first run the tests locally (some CI systems allow running tests on private changes remotely \cite{teamCityRemoteBuild}), and avoid breaking the build for the rest of the team. 
In case of a fix aimed at eliminating a bug that is not covered by existing test, we would have expected developers to add coverage, that would protect the system in case of regression.
We assume there is a CI system in place, an assumption we feel comfortable with in the context of automated testing.

\section{Semantic Changes and Test Maintenance}

\subsection{\textbf{RQ. 3.}: How do semantic changes performed in (production) code maintenance relate to test maintenance activities?}

In order to explore how test maintenance (see section \ref{sec:testRelatedOperations}) is affected by semantic change types (Fluri et al. \cite{fluri2006classifying}) in production code, we explore our dataset on an individual commit granularity. Studies show that commits which perform a large rename or move refactorings may involve a great deal of test methods and/or classes moves from one package to another \cite{pinto2012understanding}. Moreover, commits that add entire packages as part of open sourcing new modules, and other activities of similar nature can yield extreme values for test maintenance. 
Such extreme values pose a challenge to establishing reliable relationships. 
We deal with these challenges by removing commits with extreme $\operatorname{TestMaintenance}$ values by applying a technique suggested by Hubert at el. \cite{hubert2008adjusted} on \textit{positive}  $\operatorname{TestMaintenance}$ values. This technique operates similarly to the traditional IRQ method where data point greater than $Q3 + (1.5 * IQR)$ are removed, however, different thresholds are employed to accommodate skewed distributions.
In addition, since our goal is to establish relations between test maintenance and semantic change types in production code, we remove commits that deal purely with test code and have no changes outside the scope of test classes (files), we also ignore semantic change types made to test files.
The clean up stage eliminated roughly 10\% of the dataset.

To establish relationships between test maintenance and semantic change types in production code, we devise generalised logistic regressions of the form: $${\operatorname{HasK} = \operatorname{Const^{K}} + \sum\limits_{i=1}^{|\operatorname{Predictors}|}c_i^{K} * \operatorname{SCT}_i}$$ where $\operatorname{HasK}$ is a binary version of the underlying metric $K$ indicated in section \ref{sec:testRelatedOperations}, $\operatorname{SCT}_i$ is the semantic change type, $c_i^K$ are the predictor coefficients and $\operatorname{Const^K}$ is the model constant.
$\operatorname{HasK}$ evaluates to $\operatorname{TRUE}$ for a commit iff the underlying metric $K$ is greater than 0 for that particular commit.
For example, the metric: 
\begin{equation*}
\resizebox{1\linewidth}{!}{$%
\operatorname{HasTestMaintenance}(commit) := \operatorname{TestMaintenance}(commit) > 0
$%
}%
\end{equation*}
is a binary version of the $\operatorname{TestMaintenance}$ metric.
We use binary metrics to detect the presence of test maintenance, rather then quantifying it. Our goal is to determine whether a developer added any test methods, rather than detecting that a developer added exactly 5 new test methods.

We devised a regression model for $\operatorname{HasTestMaintenance}(commit)$ (omitted for brevity), which indicated that most semantic change types are statistically significant.
We then proceeded to analysing the effect of semantic changes on test maintenance activities by computing the odds ratio for each semantic change type. Semantic change types that increase the odds of the outcome being $\operatorname{TRUE}$, are greater than 1, while semantic change types that decrease the odds of the outcome being $\operatorname{TRUE}$, are less than 1.

\begin{figure}[ht]    
    %\captionsetup{belowskip=12pt}
    \caption{Odds ratio for the $\operatorname{HasTestMaintenance}$ metric}
    \label{fig:oddsRatioHasTestMaintenance}    
    \includegraphics[width=\linewidth]{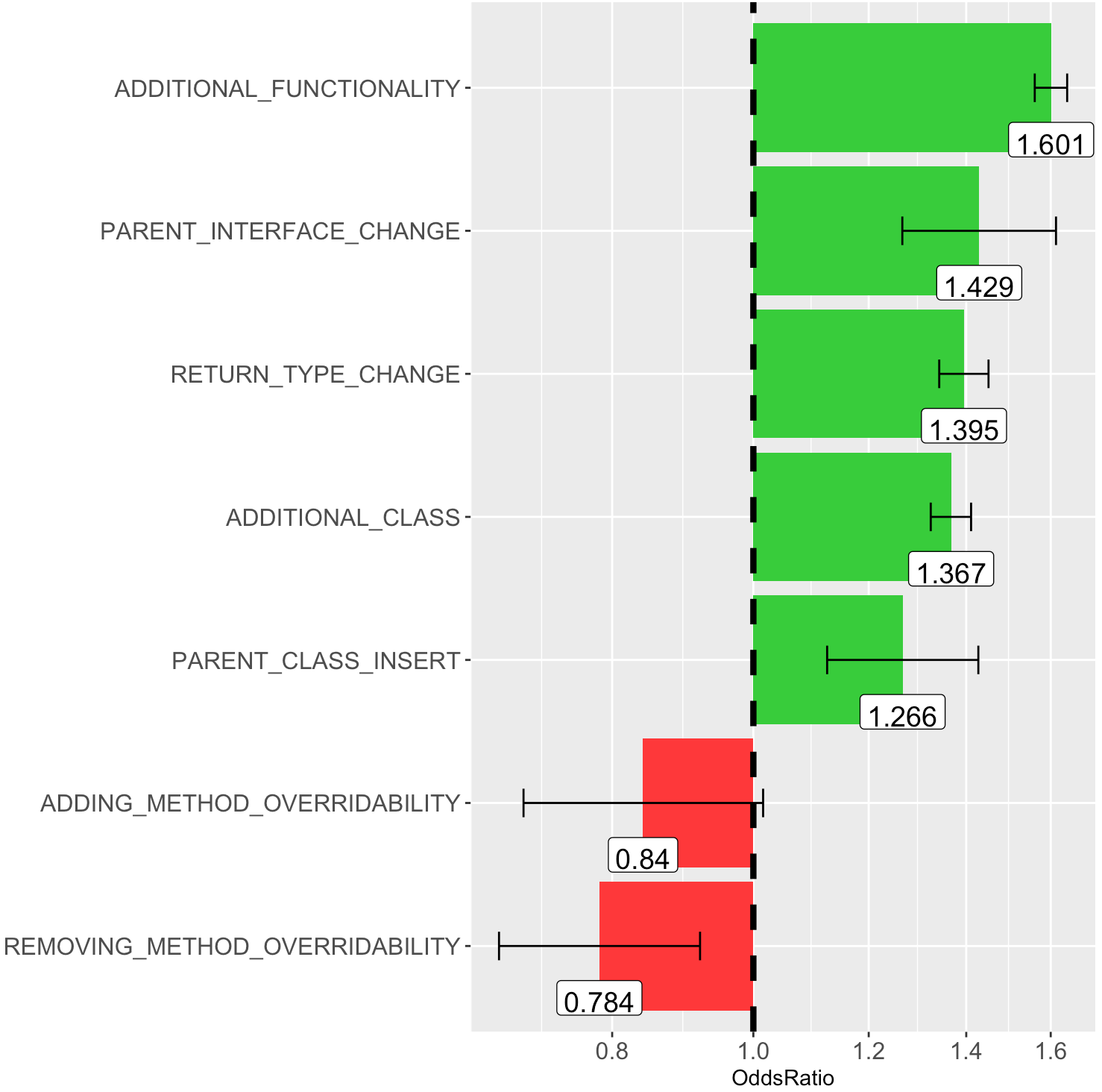}   
    %\vspace{1em}%
\end{figure}

We visualise the semantic change types with the most dominant odds ratio in figure \ref{fig:oddsRatioHasTestMaintenance}. We depict only the odds ratio of the 5 most positively, and 5 most negatively affecting predictors, and only plot predictors that demonstrate sufficient strength which we set to $|\operatorname{oddsRatio(predictor)}-1| >$ 0.15.
Green bars indicate positive effect, while red bars indicate negative effect. The 95\% confidence interval is indicated on top of every bar.

\subsection{Analysing the odds ratio}
\label{sec:rq3Discuss}

The semantic change type {\small{``ADDITIONAL\_FUNCTIONALITY''}} has odds ratio of 1.601 which means that an increase of one unit of measure in {\small{``ADDITIONAL\_FUNCTIONALITY''}} will increase the odds of $\operatorname{HasTestMaintenance}$ being $\operatorname{TRUE}$ by a factor of 1.601. In other words, with each addition of a new method to production code, the odds of test maintenance to be present in that commit increase by a factor of 1.601. \\
The odds ratio of {\small{REMOVING\_METHOD\_OVERRIDABILITY}} is 0.784, which indicates that each introduction of the semantic change type {\small{REMOVING\_METHOD\_OVERRIDABILITY}}, which stands for adding the ``final'' keyword to a method's declaration, will increase the odds of $\operatorname{HasTestMaintenance}$ being $\operatorname{TRUE}$ by a factor of 0.784, and since this factor is less than 1, it actually \textit{decreases} the odds of test maintenance to be present by a factor of $1/0.784=1.275$. In the context of odds ratio, a value $\operatorname{Val}$ such that $\operatorname{Val} < 1$ indicates a decrease by a factor of $\frac{1}{\operatorname{Val}}$ in the odds.

\begin{figure*}
    \centering
    \begin{adjustbox}{minipage=\linewidth,scale=0.95}    
    \caption{Odds ratio for test maintenance activities}
    \label{fig:oddsRatioAll}
    
    \begin{subfigure}[b]{0.45\textwidth}
        \subcaption{Odds ratio for the $\operatorname{Test^{Method}_{A}}$ metric}      
        \includegraphics[width=\linewidth]{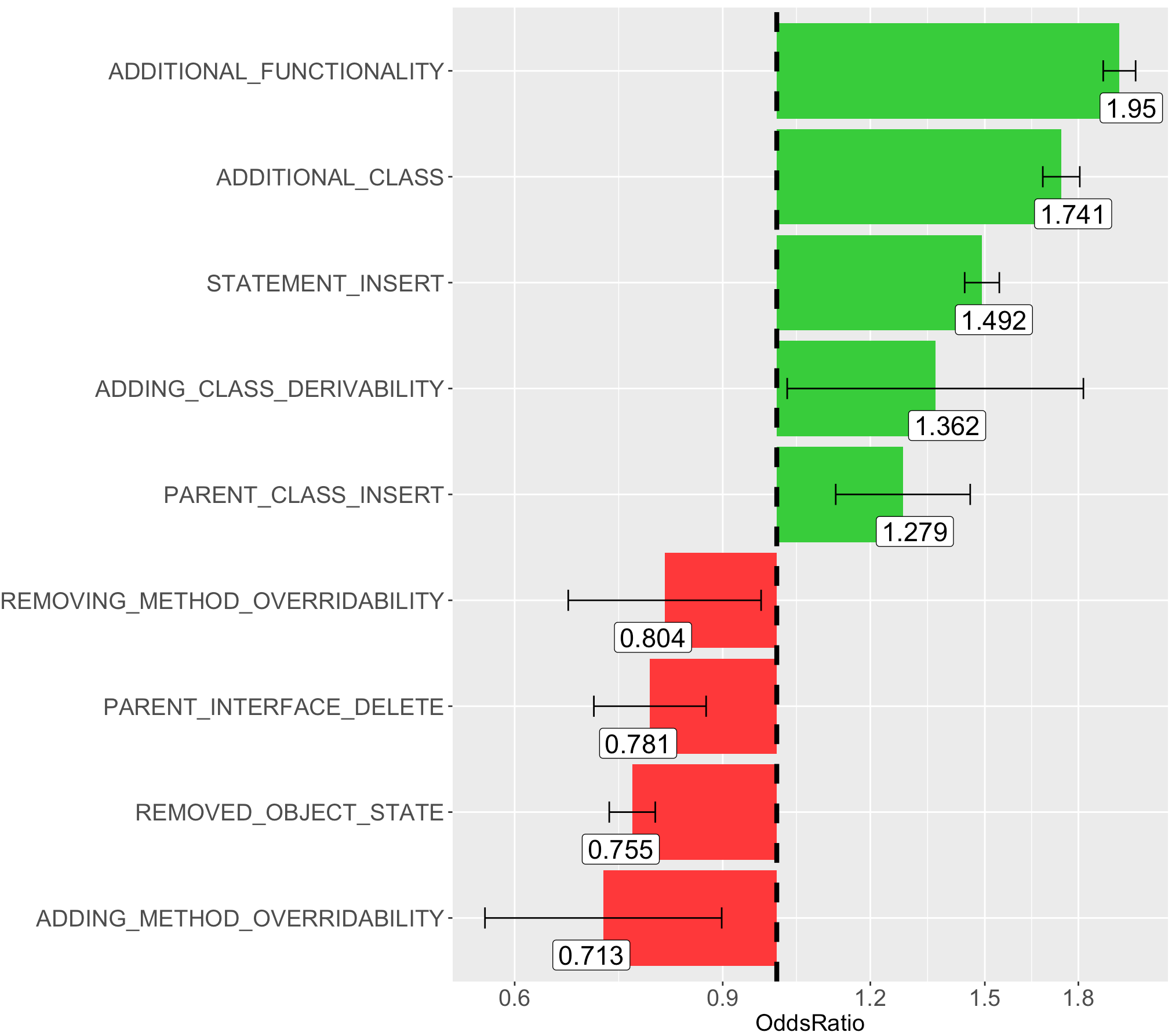}
         \label{fig:oddsRatioTestMethodAdded}
    \end{subfigure}%
    \quad
    \begin{subfigure}[b]{0.45\textwidth}
         \subcaption{Odds ratio for $\operatorname{Test^{Method}_{R}}$ metric}   
       \includegraphics[width=\linewidth]{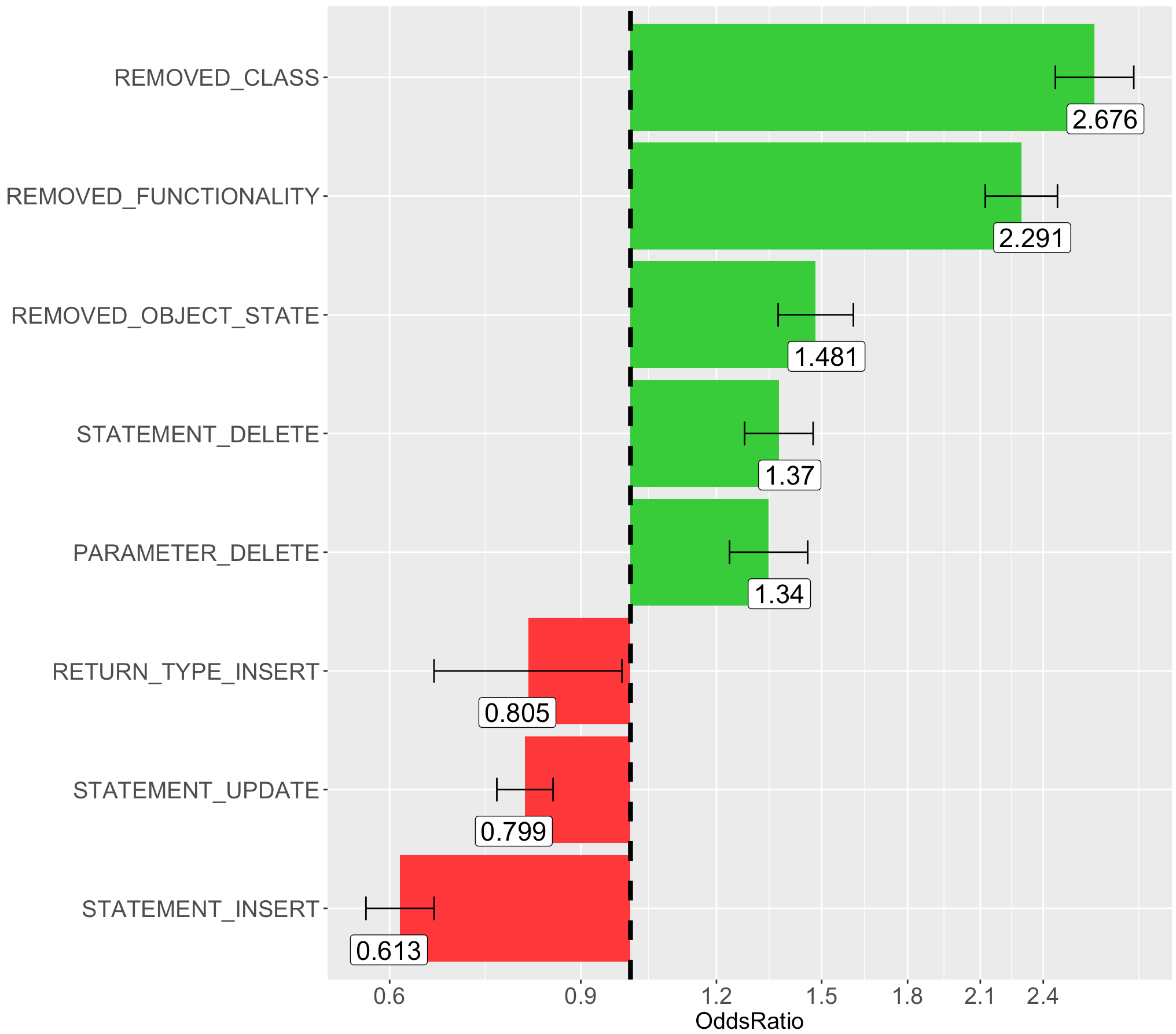}
        \label{fig:oddsRatioTestMethodRemoved}
    \end{subfigure}
 
    \begin{subfigure}[b]{0.45\textwidth}
        %\captionsetup{belowskip=12pt}
        \subcaption{Odds ratio for $\operatorname{Test^{Method}_{U}}$ metric}    
        \includegraphics[width=\linewidth]{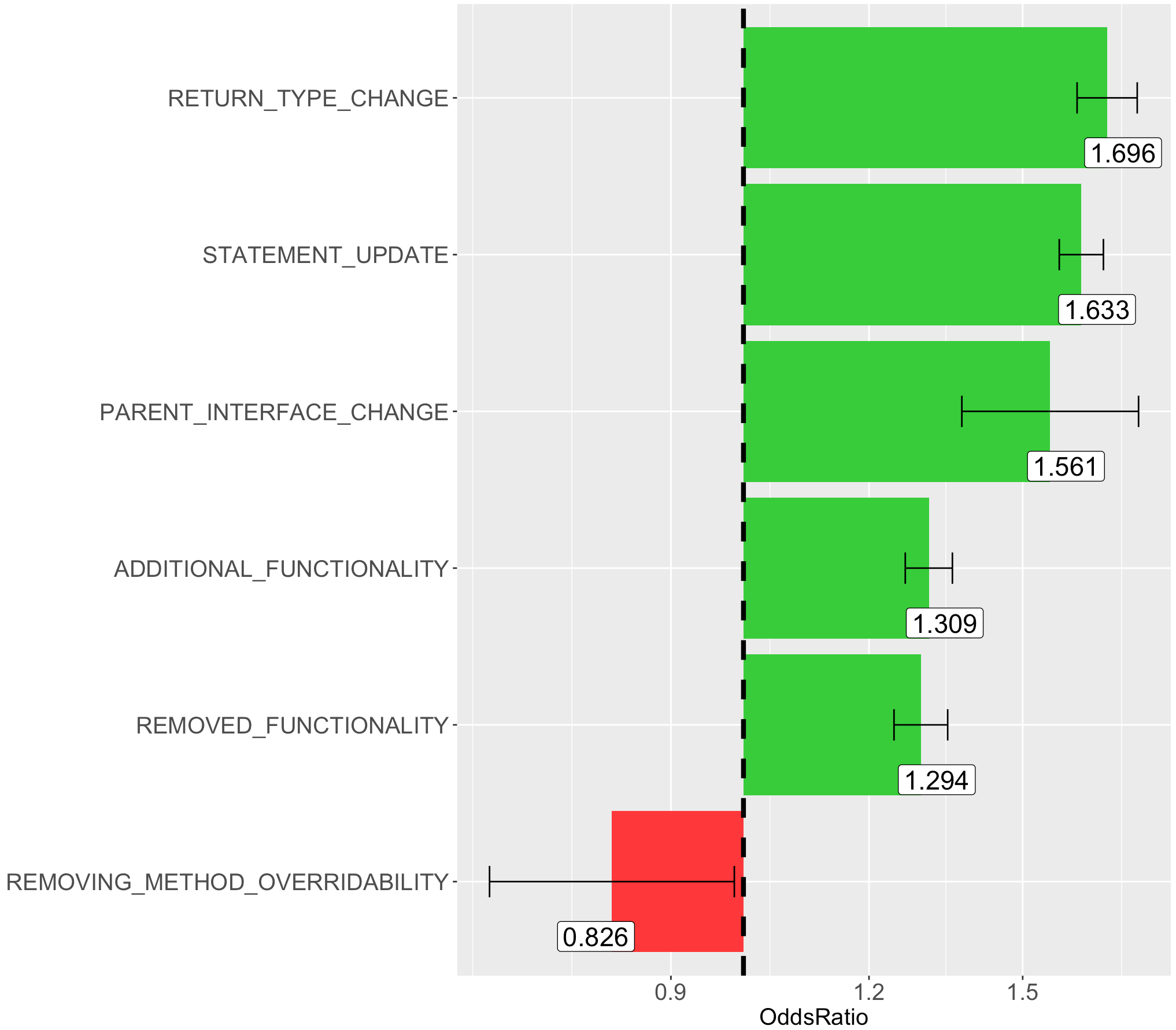}   
        \label{fig:oddsRatioTestMethodChanged}    
    \end{subfigure}
    \quad
     \begin{subfigure}[b]{0.45\textwidth}
        %\captionsetup{belowskip=12pt}
        \subcaption{Odds ratio for $\operatorname{Test^{Class}_{A}}$ metric}  
        \includegraphics[width=\linewidth]{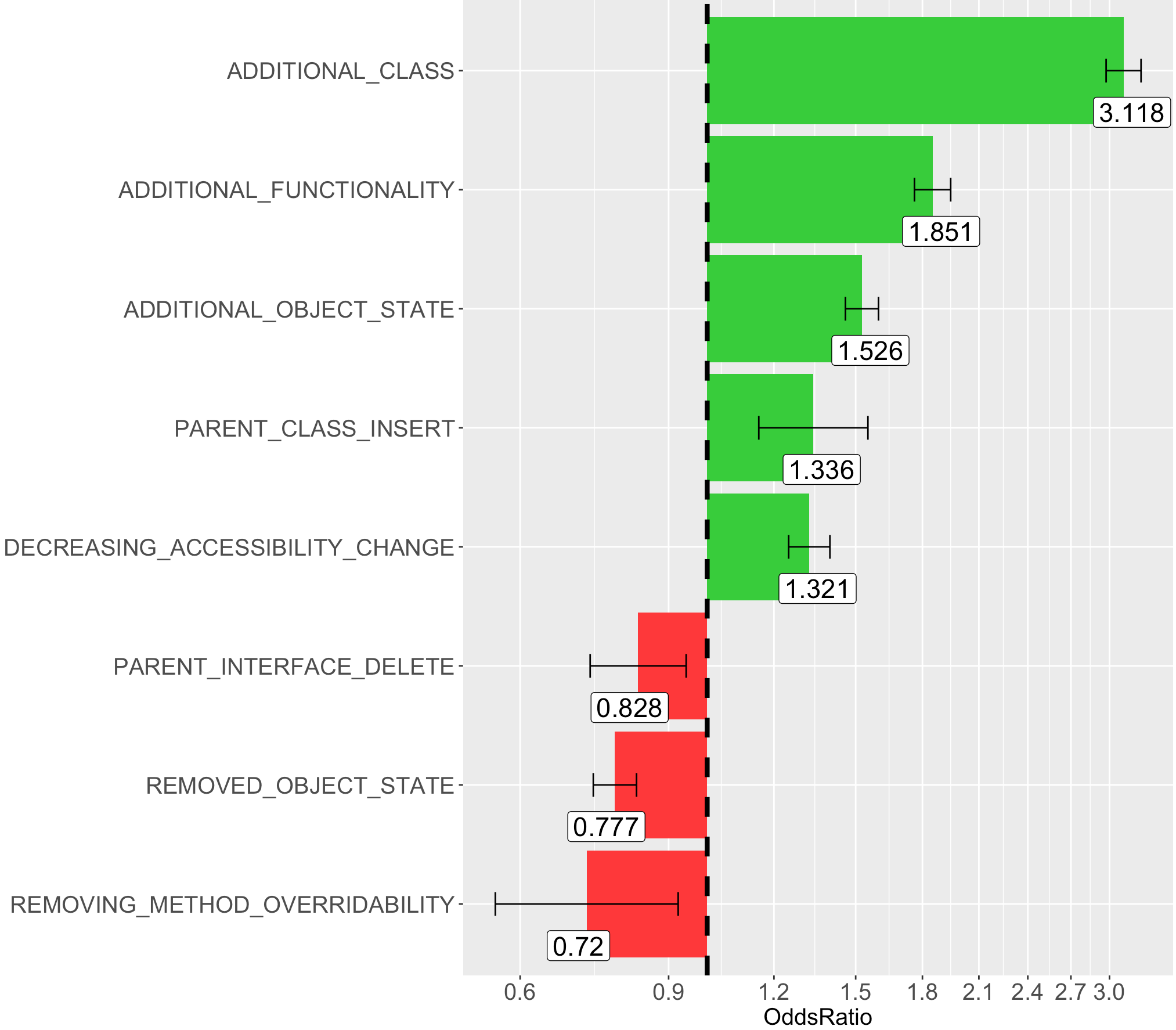}  
        \label{fig:oddsRatioTestClassAdded}    
    \end{subfigure}
    
    \begin{subfigure}[b]{0.45\textwidth}
        %\captionsetup{belowskip=12pt}
        \subcaption{Odds ratio for $\operatorname{Test^{Class}_{R}}$ metric}
        \includegraphics[width=\linewidth]{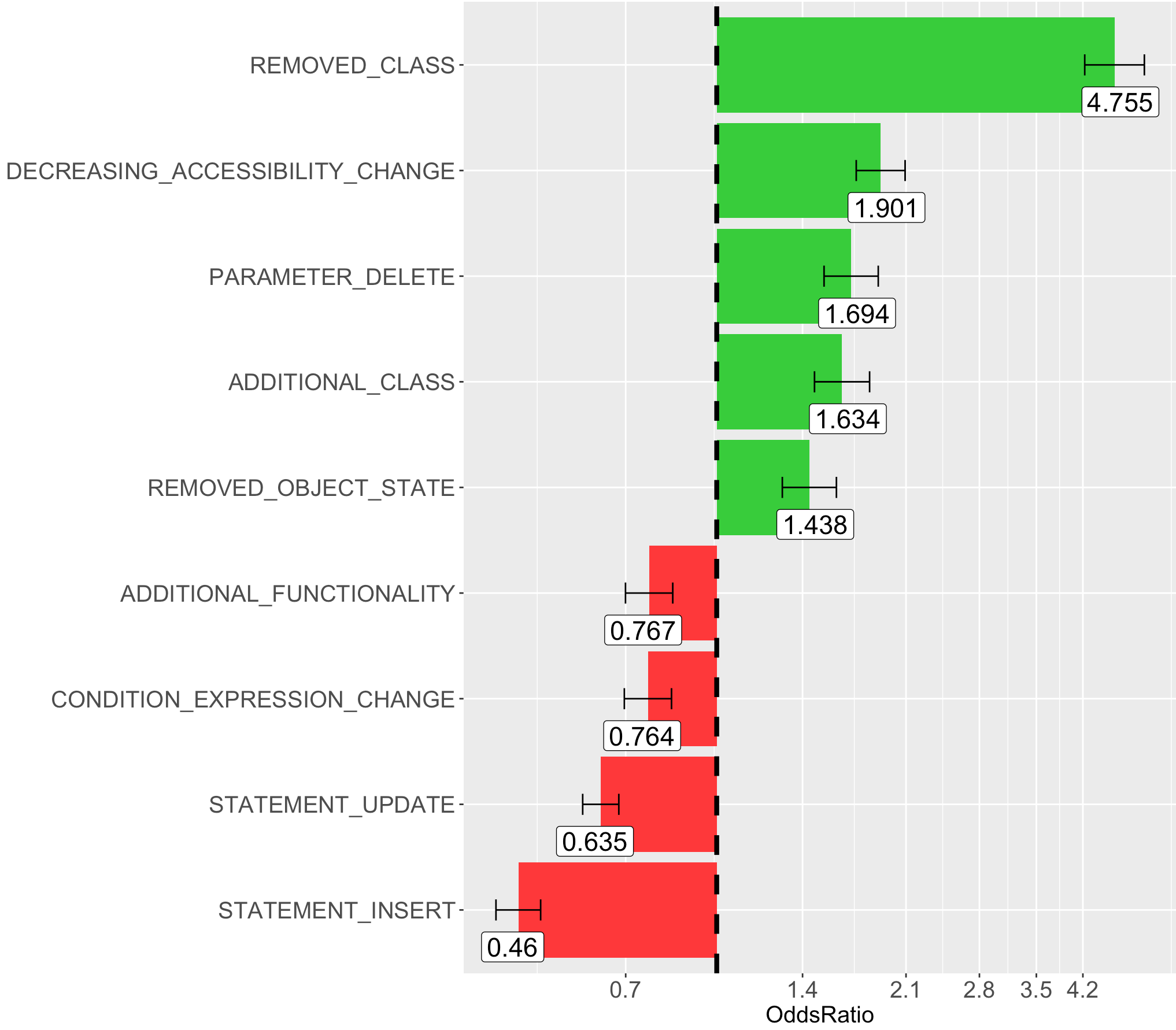} \label{fig:oddsRatioTestClassRemoved}   
    \end{subfigure}
    \quad
    \begin{subfigure}[b]{0.45\textwidth}
        %\captionsetup{belowskip=12pt}
        \subcaption{Odds ratio for $\operatorname{Test^{Class}_{U}}$ metric}        
        \includegraphics[width=\linewidth]{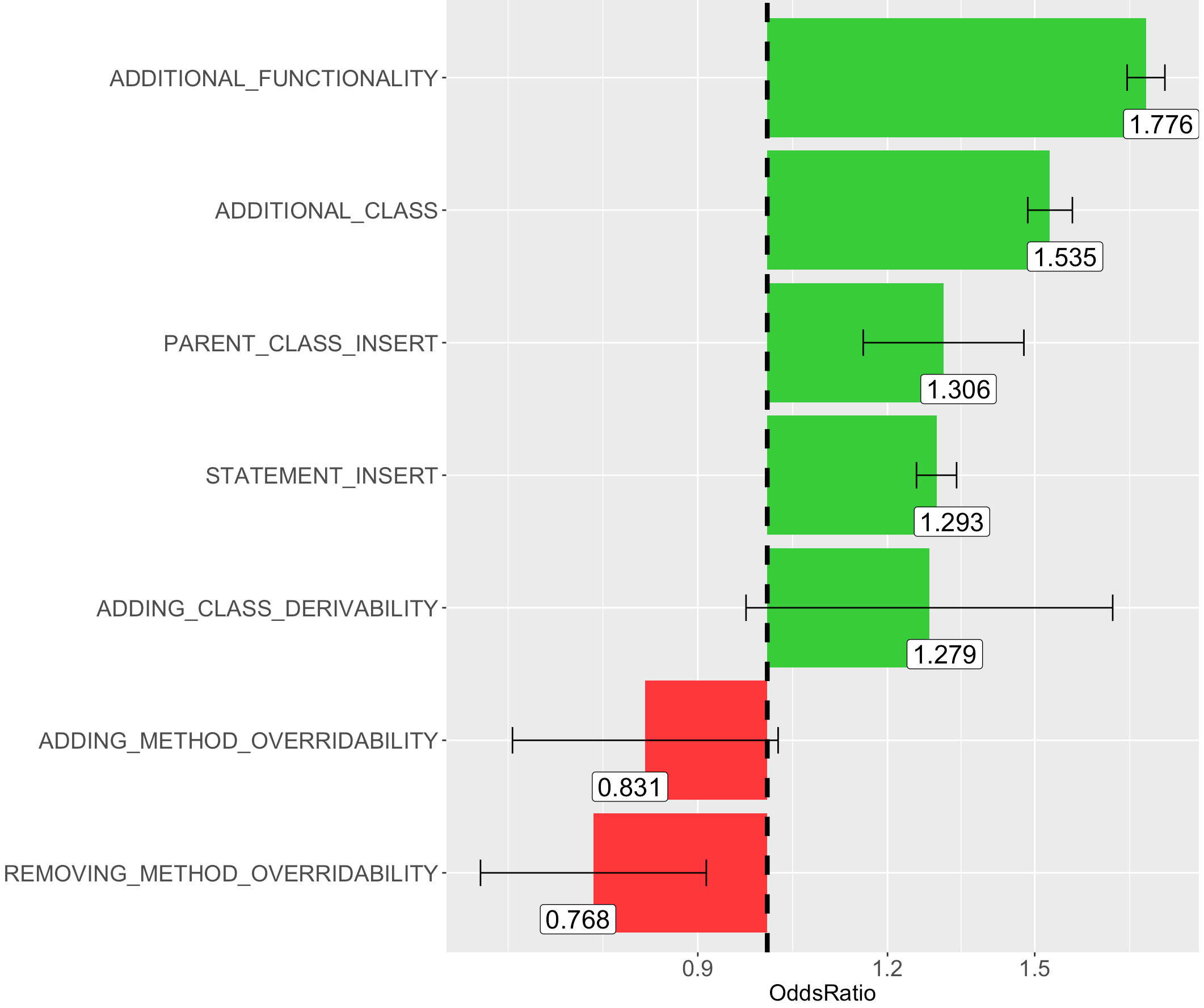}
        \label{fig:oddsRatioTestClassChanged}    
    \end{subfigure}
    
     \end{adjustbox}     
\end{figure*}

Figure \ref{fig:oddsRatioAll} visualises the odds ratio for 
$\operatorname{Test^{Method}_{A}}$, $\operatorname{Test^{Method}_{R}}$, $\operatorname{Test^{Method}_{U}}$, $\operatorname{Test^{Class}_{A}}$, $\operatorname{Test^{Class}_{R}}$, $\operatorname{Test^{Class}_{U}}$.
It can be seen that adding new methods and/or classes significantly increases the odds of a test method or a test class to be added as part of a commit, while the removal of a class field tends to decrease them (figures \ref{fig:oddsRatioTestMethodAdded},\ref{fig:oddsRatioTestClassAdded}).
Removing a class increases the odds of a test method and a test class to be removed by a factor of 2.5 and 4.7 respectively (figures \ref{fig:oddsRatioTestClassRemoved}, \ref{fig:oddsRatioTestMethodRemoved}).
Changing the return type of a method increases the odds for a test method to be updated by a factor of 1.7 (figure \ref{fig:oddsRatioTestMethodChanged}).
Adding a new statement, or updating an existing statement, significantly decreases the odds of a test class or a test method to be removed (figures \ref{fig:oddsRatioTestMethodRemoved}, \ref{fig:oddsRatioTestClassRemoved}).
Updating an existing statement significantly increases the odds of a test class to be updated (figure \ref{fig:oddsRatioTestClassChanged}).

\section{Threats to validity}
\label{sec:threatsToValidity}

\noindent\textbf{\textit{Threats to Statistical Conclusion Validity}} are the degree to which conclusions about the relationship among variables based on the data are reasonable. 
\begin{itemize}
    \item \underline{Regression Models}. Our dataset consists of 61 projects and over 240,000 commits. Both the model coefficients and the predictions were annotated with statistical significance levels to indicate the strength of the signal. Most of the coefficients were statistically significant (${\operatorname{p-value} < 0.01}$).
    To compare distributions we used the Wilcoxon-Mann-Whitney test and reported its high significance level (${\operatorname{p-value} < 0.01}$).\\
    We assume commits are independent, however, it may be the case that commits performed by the same developer share common properties.
\end{itemize}

\noindent\textbf{\textit{Threats to Construct Validity}} consider the relationship between theory and observation, in case the measured variables do not measure the actual factors.

\begin{itemize}
    \item \underline{Maintenance Activity Classification}. We employ a technique we suggested in \cite{levinICSME2017_1}, which demonstrated an accuracy of over 76\% and Cohen’s kappa over 0.63 for their test set. It may be the case that some of the commits were misclassified and introduced a bias into our dataset. The classification categories we used are widely accepted and were first put forth by Mockus \cite{mockus2000identifying}.
    Hattori et al. \cite{hattori2008nature} suggested different classification categories for maintenance activities in the context of open source projects, we did not consider these categories in the scope of this work.
    \item \underline{Semantic Change Classification}. ChangeDistiller and the VCS mining platform we have built and used are both software components, and as such, are not immune to bugs which could result in inaccurate or incomplete semantic change extraction and data aggregations.
    \item \underline{Test Maintenance Classification}. We used a widely practiced conventions and heuristics \cite{mavenTests,zaidman2011studying} for detecting JUnit test methods and test classes. However, the use of heuristics may lead to undetected test maintenance.
    \item \underline{Data Cleaning}. Prior to devising regression models, we removed extreme data points using a technique suggested in \cite{hubert2008adjusted}. 
    Despite the fact we removed only $\sim$10\% of the data, this process could have introduced bias into the dataset we operated on.
\end{itemize}

\noindent\textbf{\textit{Threats to External Validity}} consider the generalization of our findings.
\begin{itemize}
    \item \underline{Programming Language Bias}. All analyzed commits were in the Java programming language. It is possible that developers who use other programming languages, have different maintenance activity patterns which have not been explored in the scope of this work.
    \item \underline{Open Source Bias}. The repositories studied in this paper were all popular open source projects from GitHub (\cite{gitHub}). It is possible that developers' maintenance activity profiles are different in an open source environment when compared to other environments.
    \item \underline{Popularity Bias}. We intentionally selected the popular, data rich repositories. This could limit our results to developers and repositories of high popularity, and potentially skew the perspective on characteristics found only in less popular repositories and their developers.
    \item \underline{Mixed Commits}. Recent studies \cite{nguyen2013filtering,kirinuki2014hey} report that commits may involve more than one type of maintenance activity, e.g. a commit that both fixes a bug, and adds a new feature. 
    Our classification method does not currently account for such cases, but this is definitely an interesting direction to be  considered for future work (see section \ref{sec:conclusions}).
    \item \underline{Activity Boundary}. In this work we assume a commit serves as a logical boundary of an activity. It may be the case, that developers perform test maintenance as part of activities that span multiple commits. Such work patterns were not considered in the scope of this work, but are definitely an interesting direction for future work in this area.
\end{itemize}

\section{Discussion and Applications}\
\label{sec:applications}

A large body of knowledge has formed around two different aspects of defect prediction: the relationship between software defects and product metrics, and the impact of the software process metrics on the defectiveness of software. 
Existing predictive models usually employ one of the following:
\begin{description} 
    \item \textit{Code/Product metrics} - relate to the nature of the code (\cite{halstead1977elements,mccabe1976complexity}), such as lines of code (LOC), static code complexity (\cite{menzies2007data}), etc.
    \item \textit{Process metrics} - relate to the process of code change (\cite{bell2011does}), such as code churn (\cite{nagappan2005use}), change size, change complexity (\cite{hassan2009predicting}), etc.
    \item \textit{Combined metrics} - mixing process and product metrics (\cite{ostrand2011predicting, bell2013limited}).
\end{description}
Test maintenance models may complement existing models based on process and product metrics.
Integrating our test maintenance models with defect predictive models based on commit classification (e.g., Kim et al. \cite{kim2008classifying,kim2007predicting}) can be beneficial as well. For instance, our models indicate that some semantic changes considerably increase or decrease the odds for testing maintenance activities to be part of a commit. This information may be useful for predicting whether a given commit is buggy.

\section{Related Work}

Marsavina et al. \cite{marsavina2014studying} studied the co-evolution patterns
of production and test code by utilizing source code changes (which in this work we referred to as ``semantic changes'' so as to stress their structural meaning, as opposed to merely being textual edits) on 5 open source projects. They showed that several patterns were apparent, for example, upon the deletion of a production class, its associated test class is also removed, and when a new production class is added, an associated test class is also created, etc. 
In contrast to Marsavina et al., our work concentrates on exploring the relations between \textit{software maintenance activities} (see section \ref{sec:maintActivities}) and test maintenance (see section \ref{sec:testActivities}).

Zaidman et al. studied the co-evolution of test and production code \cite{zaidman2011studying} and observed two main testing strategies:
\begin{enumerate*}
    \item synchronous, where production and test code are developed simultaneously and
    \item phased, where production and test code are developed in different phases.
\end{enumerate*}
 
Pinto et al. \cite{pinto2012understanding} studied the evolution of tests between subsequent version releases and  suggested a classification for the reasons behind test addition, removal and modification. 
Moreover, their work suggested that many tests are not really deleted and added, but rather moved or renamed. Also, tests are not only added to check bug fixes or test new functionality, but also to validate changes in the code.

Fluri et al. \cite{fluri2009analyzing} analyzed the co-evolution of comments and code, and reported that the type of a source code entity, such as a method declaration or an if-statement, has a significant influence on whether or not it gets commented.

Gall et al. \cite{gall2009change} analyzed which semantic change types frequently appear together, and revealed several patterns in the way developers deal with exception flow handling, apply the single-exit principle (multiple vs. single return statements) and swap conditions in if-else statements. 

% Giger et al. \cite{giger2011comparing} compared semantic changes and code churn in the context of bug prediction. Their findings showed that semantic change types have a significantly stronger correlation with the number of bugs than a line based code churn.

Beller et al. conducted a large-scale field study, where 416 software engineers were closely monitored over the course of five months \cite{beller2015and}. Their findings indicate that software developers spend a quarter of their work time engineering tests, whereas they think they test half of their time.

Our previous work \cite{levinIcsme2016} explored semantic change types in the context of maintenance activities, and reported predictive models for developers' maintenance activity profiles. Our recent work \cite{levinICSME2017_1} also shows that semantic change types can be effectively used in combination with word frequency analysis in order to obtain high quality predictive models (accuracy and Kappa as high as 76\% and 0.63 respectively) for classifying commits into maintenance activities.

In contrast to prior work, in addition to exploring test maintenance characteristics of individual commits, we also study developers' test maintenance behaviour by considering all commits performed by the same developer.
To the best of our knowledge, our work is the first to study commit and developer level test maintenance on such scale.

\section{Conclusions and Future Work}
\label{sec:conclusions}

In this work we explore the relationships between test maintenance activities, production code maintenance activities, and semantic changes performed as part of developers' commits.\\
Our large scale empirical study provides several observations:
\begin{itemize}
    \item The number of test methods and test classes in a software project can be well predicted using models that employ code maintenance activity profiles (corrective, adaptive, perfective).
    \item Both the number of test methods and test classes has a negative correlation with corrective commits (see discussion in section \ref{sec:rq1Discussion}).
    \item Test maintenance vary considerably between projects, which may imply that current testing practices are more project specific than standardized.
    \item Empirical evidence show that among the three maintenance activity types, corrective, perfective, and adaptive, the corrective maintenance which deals with fault fixing, involve less test maintenance. We discuss and suggest several hypotheses that may account for such a result (see section \ref{sec:rq2Discuss}).
    \item Different semantic changes affect test maintenance activities differently. Moreover, certain semantic changes considerably increase or decrease the odds of a particular maintenance activity to take place in a scope of a given commit.    
\end{itemize}

% Only 1\% of the commits involve at least one test class removal, while 21\% of commits involve at least one test method update.

% We also show that half the developers in our dataset performed test maintenance in less than 14\% of their corrective commits. 

The analysis carried out in this work shows what types of changes increase and decrease the odds for test maintenance to take place. For example, with each new method added as part of a commit, the odds of it to involve a test maintenance activity increase by 60\%, while each addition of the ``final'' keyword to a method's declaration decreases such odds by 22\% (see section \ref{sec:rq2Discuss}).

We believe these insights can lead to a better understanding of software quality and help practitioners reduce costs and improve software quality.
In particular, we believe that the methods used in this work can assist in understanding what kinds of changes are usually tested, in contrast to those that are not. 
One possible direction would be integrating test maintenance models with defect prediction and test selection models, and exploring whether such a combination could improve existing results.

Future directions may also include modeling sets of semantic change in the context of test maintenance. Our intuition is that certain change types are likely to appear together, or better yet, some changes are unlikely to be made on their own, e.g, the addition of the ``final'' (which decreases the odds for test maintenance to occur by a factor of 1.275) word to a method is likely to be part of a larger change, not a commit on its own.

Activity boundary is another direction that may be worth exploring in the context of test maintenance activities. In this work we set the commit as an activity boundary, but it is also reasonable to assume some activities span multiple commits. Therefore, it can be beneficial to explore test maintenance activities in the scope of these multi-commit activities, and explore whether certain developers tend to perform code maintenance and test maintenance in separate commits, but as a single logical activity. Moreover, exploring if an activity boundary may cross the scope of a single developer may also be of interest.
Consider the scenario of multiple developers working on a single logical activity, where one or more of them perform test maintenance in designated commits, separated form the non-test code.

It could also be beneficial to further investigate the negative correlation of corrective maintenance and the test method and class count. Specifically, the question of whether test methods (and/or classes) and corrective maintenance have a cause-and-effect relationship remains open.

\clearpage

\bibliographystyle{IEEEtran}
\bibliography{bibliography}

\ifx\undefined\allcaps\def\allcaps#1{#1}\fi
\begin{thebibliography}{10}
\providecommand{\url}[1]{#1}
\csname url@samestyle\endcsname
\providecommand{\newblock}{\relax}
\providecommand{\bibinfo}[2]{#2}
\providecommand{\BIBentrySTDinterwordspacing}{\spaceskip=0pt\relax}
\providecommand{\BIBentryALTinterwordstretchfactor}{4}
\providecommand{\BIBentryALTinterwordspacing}{\spaceskip=\fontdimen2\font plus
\BIBentryALTinterwordstretchfactor\fontdimen3\font minus
  \fontdimen4\font\relax}
\providecommand{\BIBforeignlanguage}[2]{{%
\expandafter\ifx\csname l@#1\endcsname\relax
\typeout{** WARNING: IEEEtran.bst: No hyphenation pattern has been}%
\typeout{** loaded for the language `#1'. Using the pattern for}%
\typeout{** the default language instead.}%
\else
\language=\csname l@#1\endcsname
\fi
#2}}
\providecommand{\BIBdecl}{\relax}
\BIBdecl

\bibitem{hamill2004unit}
P.~Hamill, \emph{Unit Test Frameworks: Tools for High-Quality Software
  Development}.\hskip 1em plus 0.5em minus 0.4em\relax " O'Reilly Media, Inc.",
  2004.

\bibitem{levinIcsme2016}
S.~Levin and A.~Yehudai, ``Using temporal and semantic developer-level
  information to predict maintenance activity profiles,'' in \emph{Proc.\
  ICSME}.\hskip 1em plus 0.5em minus 0.4em\relax IEEE, 2016, pp. 463--468.

\bibitem{levinICSME2017_1}
S.~Levin and A.~Yehudai, ``Boosting automatic commit classification into
  maintenance activities by utilizing source code changes.''\hskip 1em plus
  0.5em minus 0.4em\relax Submitted for publication, 2017.

\bibitem{fluri2007change}
B.~Fluri, M.~Wursch, M.~PInzger, and H.~C. Gall, ``Change distilling: Tree
  differencing for fine-grained source code change extraction,'' \emph{Software
  Engineering, IEEE Transactions on}, vol.~33, no.~11, pp. 725--743, 2007.

\bibitem{marsavina2014studying}
C.~Marsavina, D.~Romano, and A.~Zaidman, ``Studying fine-grained co-evolution
  patterns of production and test code,'' in \emph{Source Code Analysis and
  Manipulation (SCAM), 2014 IEEE 14th International Working Conference
  on}.\hskip 1em plus 0.5em minus 0.4em\relax IEEE, 2014, pp. 195--204.

\bibitem{mockus2000identifying}
A.~Mockus and L.~G. Votta, ``Identifying reasons for software changes using
  historic databases,'' in \emph{Software Maintenance, 2000. Proceedings.
  International Conference on}.\hskip 1em plus 0.5em minus 0.4em\relax IEEE,
  2000, pp. 120--130.

\bibitem{lehman2000evolution}
M.~M. Lehman, J.~F. Ramil, and G.~Kahen, ``Evolution as a noun and evolution as
  a verb,'' in \emph{SOCE 2000 Workshop on Software and Organisation
  Co-evolution}, vol.~9, 2000, p.~31.

\bibitem{lehman2003software}
M.~M. Lehman and J.~F. Ramil, ``Software evolution—background, theory,
  practice,'' \emph{Information Processing Letters}, vol.~88, no.~1, pp.
  33--44, 2003.

\bibitem{gitHub}
``Github - the largest open source community in the world,''
  \url{https://github.com/open-source}, [Online; accessed 18-April-2016].

\bibitem{bitbucket}
``Bitbucket - code, manage, collaborate,'' \url{https://bitbucket.org/},
  [Online; accessed 18-April-2016].

\bibitem{sourceForge}
``Sourceforge - open source community resource dedicated to helping open source
  projects,'' \url{https://sourceforge.net/}, [Online; accessed 18-April-2016].

\bibitem{codeplex}
``Codeplex - code talks let your voice be heard,''
  \url{https://www.codeplex.com/}, [Online; accessed 18-April-2016].

\bibitem{googleCode}
``Google code,'' \url{https://code.google.com/archive/d/code.google.com},
  [Online; accessed 18-April-2016].

\bibitem{diebold2012origin}
F.~X. Diebold, ``On the origin (s) and development of the term {Big Data},''
  PIER Working Paper, 2012.

\bibitem{raychev2015predicting}
V.~Raychev, M.~Vechev, and A.~Krause, ``Predicting program properties from big
  code,'' in \emph{ACM SIGPLAN Notices}, vol.~50, no.~1.\hskip 1em plus 0.5em
  minus 0.4em\relax ACM, 2015, pp. 111--124.

\bibitem{zaharia2010spark}
M.~Zaharia, M.~Chowdhury, M.~J. Franklin, S.~Shenker, and I.~Stoica, ``Spark:
  Cluster computing with working sets.'' \emph{HotCloud}, vol.~10, pp. 10--10,
  2010.

\bibitem{sparkSite}
``Lightning-fast cluster computing,'' \url{http://spark.apache.org/}, [Online;
  accessed 11-April-2016].

\bibitem{gall2009change}
H.~C. Gall, B.~Fluri, and M.~Pinzger, ``Change analysis with evolizer and
  changedistiller,'' \emph{IEEE Software}, vol.~26, no.~1, p.~26, 2009.

\bibitem{fluri2008discovering}
B.~Fluri, E.~Giger, and H.~C. Gall, ``Discovering patterns of change types,''
  in \emph{Automated Software Engineering, 2008. ASE 2008. 23rd IEEE/ACM
  International Conference on}.\hskip 1em plus 0.5em minus 0.4em\relax IEEE,
  2008, pp. 463--466.

\bibitem{martinez2013automatically}
M.~Martinez, L.~Duchien, and M.~Monperrus, ``Automatically extracting instances
  of code change patterns with ast analysis,'' \emph{arXiv preprint
  arXiv:1309.3730}, 2013.

\bibitem{ho1998random}
T.~K. Ho, ``The random subspace method for constructing decision forests,''
  \emph{IEEE transactions on pattern analysis and machine intelligence},
  vol.~20, no.~8, pp. 832--844, 1998.

\bibitem{breiman2001random}
L.~Breiman, ``Random forests,'' \emph{Machine learning}, vol.~45, no.~1, pp.
  5--32, 2001.

\bibitem{mavenTests}
``Maven surefire plugin,''
  \url{http://maven.apache.org/surefire/maven-surefire-plugin/examples/inclusion-exclusion.html},
  [Online; accessed Jan-2017].

\bibitem{zaidman2011studying}
A.~Zaidman, B.~Van~Rompaey, A.~van Deursen, and S.~Demeyer, ``Studying the
  co-evolution of production and test code in open source and industrial
  developer test processes through repository mining,'' \emph{Empirical
  Software Engineering}, vol.~16, no.~3, pp. 325--364, 2011.

\bibitem{fluri2006classifying}
B.~Fluri and H.~C. Gall, ``Classifying change types for qualifying change
  couplings,'' in \emph{Program Comprehension, 2006. ICPC 2006. 14th IEEE
  International Conference on}.\hskip 1em plus 0.5em minus 0.4em\relax IEEE,
  2006, pp. 35--45.

\bibitem{junitTestFixture}
``Test fixtures,''
  \url{https://github.com/junit-team/junit4/wiki/Test-fixtures}, [Online;
  accessed 26-March-2017].

\bibitem{mccullagh1989generalized}
P.~McCullagh and J.~A. Nelder, \emph{Generalized linear models}.\hskip 1em plus
  0.5em minus 0.4em\relax CRC press, 1989, vol.~37.

\bibitem{chambers1991statistical}
J.~M. Chambers and T.~J. Hastie, \emph{Statistical models in S}.\hskip 1em plus
  0.5em minus 0.4em\relax CRC Press, Inc., 1991.

\bibitem{fulton2012confusion}
L.~V. Fulton, F.~A. Mendez, N.~D. Bastian, and R.~M. Musal, ``Confusion between
  odds and probability, a pandemic?'' \emph{Journal of Statistics Education},
  vol.~20, no.~3, p.~n3, 2012.

\bibitem{R}
\BIBentryALTinterwordspacing
{R Development Core Team}, \emph{R: A Language and Environment for Statistical
  Computing}, R Foundation for Statistical Computing, Vienna, Austria, 2008,
  {ISBN} 3-900051-07-0. [Online]. Available: \url{http://www.R-project.org}
\BIBentrySTDinterwordspacing

\bibitem{rCaret}
``The caret package,'' \url{http://topepo.github.io/caret/index.html}, [Online;
  accessed Nov-2016].

\bibitem{shihab2012exploration}
E.~Shihab, ``An exploration of challenges limiting pragmatic software defect
  prediction,'' Ph.D. dissertation, Citeseer, 2012.

\bibitem{camargo2009towards}
A.~E. Camargo~Cruz and K.~Ochimizu, ``Towards logistic regression models for
  predicting fault-prone code across software projects,'' in \emph{Proceedings
  of the 2009 3rd International Symposium on Empirical Software Engineering and
  Measurement}.\hskip 1em plus 0.5em minus 0.4em\relax IEEE Computer Society,
  2009, pp. 460--463.

\bibitem{venables2013modern}
W.~N. Venables and B.~D. Ripley, \emph{Modern applied statistics with
  S-PLUS}.\hskip 1em plus 0.5em minus 0.4em\relax Springer Science \& Business
  Media, 2013.

\bibitem{bauer1972constructing}
D.~F. Bauer, ``Constructing confidence sets using rank statistics,''
  \emph{Journal of the American Statistical Association}, vol.~67, no. 339, pp.
  687--690, 1972.

\bibitem{fowler2006continuous}
M.~Fowler and M.~Foemmel, ``Continuous integration,'' \emph{Thought-Works)
  http://www. thoughtworks. com/Continuous Integration. pdf}, 2006.

\bibitem{jenkins}
``Build great things at any scale,'' \url{https://jenkins.io/}, [Online;
  accessed 26-March-2017].

\bibitem{teamCity}
``Powerful continuous integration out of the box,''
  \url{https://www.jetbrains.com/teamcity/}, [Online; accessed 26-March-2017].

\bibitem{teamCityRemoteBuild}
``Remote run / personal build,''
  \url{https://confluence.jetbrains.com/display/TCD10/Remote+Run}, [Online;
  accessed 26-March-2017].

\bibitem{pinto2012understanding}
L.~S. Pinto, S.~Sinha, and A.~Orso, ``Understanding myths and realities of
  test-suite evolution,'' in \emph{Proceedings of the ACM SIGSOFT 20th
  International Symposium on the Foundations of Software Engineering}.\hskip
  1em plus 0.5em minus 0.4em\relax ACM, 2012, p.~33.

\bibitem{hubert2008adjusted}
M.~Hubert and E.~Vandervieren, ``An adjusted boxplot for skewed
  distributions,'' \emph{Computational statistics \& data analysis}, vol.~52,
  no.~12, pp. 5186--5201, 2008.

\bibitem{hattori2008nature}
L.~P. Hattori and M.~Lanza, ``On the nature of commits,'' in \emph{Proceedings
  of the 23rd IEEE/ACM International Conference on Automated Software
  Engineering}.\hskip 1em plus 0.5em minus 0.4em\relax IEEE Press, 2008, pp.
  III--63.

\bibitem{nguyen2013filtering}
H.~A. Nguyen, A.~T. Nguyen, and T.~N. Nguyen, ``Filtering noise in
  mixed-purpose fixing commits to improve defect prediction and localization,''
  in \emph{Software Reliability Engineering (ISSRE), 2013 IEEE 24th
  International Symposium on}.\hskip 1em plus 0.5em minus 0.4em\relax IEEE,
  2013, pp. 138--147.

\bibitem{kirinuki2014hey}
H.~Kirinuki, Y.~Higo, K.~Hotta, and S.~Kusumoto, ``Hey! are you committing
  tangled changes?'' in \emph{Proceedings of the 22nd International Conference
  on Program Comprehension}.\hskip 1em plus 0.5em minus 0.4em\relax ACM, 2014,
  pp. 262--265.

\bibitem{halstead1977elements}
M.~H. Halstead, \emph{Elements of software science}.\hskip 1em plus 0.5em minus
  0.4em\relax Elsevier New York, 1977, vol.~7.

\bibitem{mccabe1976complexity}
T.~J. McCabe, ``A complexity measure,'' \emph{Software Engineering, IEEE
  Transactions on}, no.~4, pp. 308--320, 1976.

\bibitem{menzies2007data}
T.~Menzies, J.~Greenwald, and A.~Frank, ``Data mining static code attributes to
  learn defect predictors,'' \emph{Software Engineering, IEEE Transactions on},
  vol.~33, no.~1, pp. 2--13, 2007.

\bibitem{bell2011does}
R.~M. Bell, T.~J. Ostrand, and E.~J. Weyuker, ``Does measuring code change
  improve fault prediction?'' in \emph{Proceedings of the 7th International
  Conference on Predictive Models in Software Engineering}.\hskip 1em plus
  0.5em minus 0.4em\relax ACM, 2011, p.~2.

\bibitem{nagappan2005use}
N.~Nagappan and T.~Ball, ``Use of relative code churn measures to predict
  system defect density,'' in \emph{Software Engineering, 2005. ICSE 2005.
  Proceedings. 27th International Conference on}.\hskip 1em plus 0.5em minus
  0.4em\relax IEEE, 2005, pp. 284--292.

\bibitem{hassan2009predicting}
A.~E. Hassan, ``Predicting faults using the complexity of code changes,'' in
  \emph{Proceedings of the 31st International Conference on Software
  Engineering}.\hskip 1em plus 0.5em minus 0.4em\relax IEEE Computer Society,
  2009, pp. 78--88.

\bibitem{ostrand2011predicting}
T.~J. Ostrand and E.~J. Weyuker, ``Predicting bugs in large industrial software
  systems.'' in \emph{ISSSE}.\hskip 1em plus 0.5em minus 0.4em\relax Springer,
  2011, pp. 71--93.

\bibitem{bell2013limited}
R.~M. Bell, T.~J. Ostrand, and E.~J. Weyuker, ``The limited impact of
  individual developer data on software defect prediction,'' \emph{Empirical
  Software Engineering}, vol.~18, no.~3, pp. 478--505, 2013.

\bibitem{kim2008classifying}
S.~Kim, E.~J. Whitehead~Jr, and Y.~Zhang, ``Classifying software changes: Clean
  or buggy?'' \emph{IEEE Transactions on Software Engineering}, vol.~34, no.~2,
  pp. 181--196, 2008.

\bibitem{kim2007predicting}
S.~Kim, T.~Zimmermann, E.~J. Whitehead~Jr, and A.~Zeller, ``Predicting faults
  from cached history,'' in \emph{Proceedings of the 29th international
  conference on Software Engineering}.\hskip 1em plus 0.5em minus 0.4em\relax
  IEEE Computer Society, 2007, pp. 489--498.

\bibitem{fluri2009analyzing}
B.~Fluri, M.~W{\"u}rsch, E.~Giger, and H.~C. Gall, ``Analyzing the co-evolution
  of comments and source code,'' \emph{Software Quality Journal}, vol.~17,
  no.~4, pp. 367--394, 2009.

\bibitem{beller2015and}
M.~Beller, G.~Gousios, A.~Panichella, and A.~Zaidman, ``When, how, and why
  developers (do not) test in their ides,'' in \emph{Proceedings of the 2015
  10th Joint Meeting on Foundations of Software Engineering}.\hskip 1em plus
  0.5em minus 0.4em\relax ACM, 2015, pp. 179--190.

\end{thebibliography}

\end{document}